\documentstyle{article}

\makeatletter
\newif\ifITeX
\ITeXtrue

\makeatletter
\def\be{%
\@ifnextchar[
{\def\ee{\end{equation}}\begin{equation}\l@b}%
{\def\ee{$$}$$}%
}
\def\l@b[#1]{\label{#1}}
\def\eq#1{(\ref{#1})}
\def\Eq#1{Eq.(\ref{#1})}
\let\SUB\S
\def\sect#1{\ref{#1}}
\def\Sect#1{sect.\ref{#1}}
\def\subsect#1{\SUB\ref{#1}}
\def\cref#1{#1}
\@addtoreset{equation}{section}%
\ifITeX
\def\BUG#1{\vrule width 2pt height 8pt depth 2pt\relax
\typeout{BUG? line= \the\inputlineno: #1}}
\def\dubious{\typeout{DUBIOUS: line=\the\inputlineno}}
\else
\def\BUG#1{\vrule width 2pt height 8pt depth 2pt\relax
\typeout{BUG? page= \thepage: #1}}
\def\dubious{\typeout{DUBIOUS: page= \thepage}}
\fi
\ifITeX
\def\ref#1{\@ifundefined{r@#1}{{\bf ??}\typeout
   {! Reference `#1' undefined}\typeout{l.\the\inputlineno
   }}{\edef\@tempa{\@nameuse{r@#1}}\expandafter
    \@car\@tempa \@nil\null}}
\fi
\def\EOE#1{#1\gobblespaces}
\let\EOH\relax
\long\def\gobblespaces#1\ee{\ee}
\def\@cite#1#2{\hbox{ [#1\if@tempswa ,#2\fi]}}
\def\@citex[#1]#2{\if@filesw\immediate\write\@auxout{\string\citation{#2}}\fi
  \def\@citea{}\@cite{\@for\@citeb:=#2\do
    {\@citea\def\@citea{,\penalty\@m}\@ifundefined  
       {b@\@citeb}{{\bf ?}\@warning
       {Citation `\@citeb' on \ifITeX line \the\inputlineno
                              \else page \thepage \fi
        \space undefined}}%
\hbox{\csname b@\@citeb\endcsname}}}{#1}}
\def\old@sect#1#2#3#4#5#6[#7]#8{\ifnum #2>\c@secnumdepth
     \def\@svsec{}\else
     \refstepcounter{#1}\edef\@svsec{\csname the#1\endcsname.%
     \hskip 0.8em }\fi
     \@tempskipa #5\relax
      \ifdim \@tempskipa>\z@
        \begingroup #6\relax
          \@hangfrom{\hskip #3\relax\@svsec}{\interlinepenalty \@M #8\par}%
        \endgroup
       \csname #1mark\endcsname{#7}\addcontentsline
         {toc}{#1}{\ifnum #2>\c@secnumdepth \else
                      \protect\numberline{\csname the#1\endcsname}\fi
                    #7}\else
        \def\@svsechd{#6\hskip #3\@svsec #8\csname #1mark\endcsname
                      {#7}\addcontentsline
                           {toc}{#1}{\ifnum #2>\c@secnumdepth \else
                             \protect\numberline{\csname the#1\endcsname}\fi
                       #7}}\fi
     \@xsect{#5}}
\def\@sect#1#2#3#4#5#6[#7]#8{\ifnum #2>\c@secnumdepth
     \def\@svsec{}\else
     \refstepcounter{#1}\edef\@svsec%
{\ifnum #2=2{\SUB\kern-.43em\SUB\kern.20em}\fi%
\csname the#1\endcsname%
.
     \hskip 0.8em }\fi
     \@tempskipa #5\relax
      \ifdim \@tempskipa>\z@
        \begingroup #6\relax
          \@hangfrom{\hskip #3\relax\@svsec}{\interlinepenalty \@M #8\par}%
        \endgroup
       \csname #1mark\endcsname{#7}\addcontentsline
         {toc}{#1}{\ifnum #2>\c@secnumdepth \else
                      \protect\numberline{%
\ifnum #2=2\SUB\fi%
                      \csname the#1\endcsname}\fi
                    #7}\else
        \def\@svsechd{#6\hskip #3\@svsec #8\csname #1mark\endcsname
                      {#7}\addcontentsline
                           {toc}{#1}{\ifnum #2>\c@secnumdepth \else
                             \protect\numberline{%
\ifnum #2=2\SUB\fi%
                             \csname the#1\endcsname}\fi
                       #7}}\fi
     \@xsect{#5}}
\def\herring{\@ifnextchar[{\@herring}{\@herring[\vcenter]}}
\def\@herring[#1]#2{\begingroup
\def\*{\\ \>}
\topsep0pt
\partopsep0pt
\def\tabbing{\lineskip\jot \lineskiplimit\jot
     \let\>\@rtab\let\<\@ltab\let\=\@settab
     \let\+\@tabplus\let\-\@tabminus\let\`\@tabrj\let\'\@tablab
     \let\\=\@tabcr
     \global\@hightab\@firsttab
     \global\@nxttabmar\@firsttab
     \dimen\@firsttab\@totalleftmargin
     \global\@tabpush0 \global\@rjfieldfalse
     \trivlist \item[]\if@minipage\else\vskip\parskip\fi
     \setbox\@tabfbox\hbox{\rlap{\indent\hskip\@totalleftmargin
       \the\everypar}}\def\@itemfudge{\box\@tabfbox}\@startline\ignorespaces}
\def\@startfield{\global\setbox\@curfield\hbox
                    \bgroup$\displaystyle}%
\def\@stopfield{$\egroup}%
#1{\begin{tabbing}#2\end{tabbing}}\endgroup}
\def\thebibliography#1{\subsection*{References\@mkboth
 {REFERENCES}{REFERENCES}}\list
 {[\arabic{enumi}]}{\settowidth\labelwidth{[#1]}\leftmargin\labelwidth
 \advance\leftmargin\labelsep
 \usecounter{enumi}}
 \def\newblock{\hskip .11em plus .33em minus .07em}
 \sloppy\clubpenalty4000\widowpenalty4000
 \sfcode`\.=1000\parskip-1pt\relax}
\def\textindent#1{\leavevmode\par{\parindent3em\indent\llap{#1\enspace}}\ignorespaces}
\def\bigtextindent#1{\leavevmode\par{\parindent4em\indent\llap{#1\enspace}}\ignorespaces}
\newenvironment{appendices}{\begingroup
\let\@sect\old@sect
\section*{Appendices\EOH.}%
\setcounter{subsection}{0}
\@addtoreset{equation}{subsection}%
\def\appendix##1{\subsection{##1}}%
}{\endgroup}

\let\tcap\cap
\let\tcup\cup
\def\cap{\mathop{\tcap}}
\def\cup{\mathop{\tcup}}
\def\tAs{{\it As\/}}
\def\tIR{{\rm IR}}
\def\tUV{{\rm UV}}
\def\IR{{\it IR}}
\def\UV{{\it UV}}
\def\E{\tilde{{\bf E}}}
\def\W{W}
\def\Z{Z}
\def\M{M}
\def\0{{\protect\vphantom{0}}}
\def\Cut{\Phi}
\def\cut{\phi}
\def\G{{\mathchoice{G}{{\textstyle G}}{{\scriptscriptstyle G}}{{\scriptscriptstyle G}}}}
\def\Gam{{\mathchoice{\Gamma}{{\textstyle \Gamma}}{{\scriptscriptstyle \Gamma}}{{\scriptscriptstyle \Gamma}}}}
\def\gam{{\mathchoice{\gamma}{{\textstyle \gamma}}{{\scriptscriptstyle \gamma}}{{\scriptscriptstyle \gamma}}}}
\def\GG{{\G\mbackslash\Gam}}
\def\mbackslash{{\mathchoice{\backslash}{{\textstyle\backslash}}{{\scriptscriptstyle\backslash}}{{\scriptscriptstyle\backslash}}}}
\def\mxi{{\mathchoice{\xi}{{\textstyle\xi}}{{\scriptscriptstyle\xi}}{{\scriptscriptstyle\xi}}}}
\def\Gg{{\G\mbackslash\gam}}
\def\Gx{{\G\mbackslash\mxi}}
\def\equals{\mathop{=}}
\def\bydef{\mathrel{{\displaystyle\equals^{\rm def}}}}
\def\prec{<}
\def\succ{>}

\def \r   { \tilde{\bf r}   }
\def \cR {{ \cal R          }}
\def \R   { \tilde{\bf R}   }
\def \Rm  { \R^m            }
\def \As {{ \bf As          }}
\def \Asm { \As_m           }
\def \Ask { \As_\kappa      }
\def \Ass { \As_\sigma      }
\def \Asks{ \As_{\k,\s}     }
\def \Aspk{ \As'_\kappa     }
\def \rAs { \r\.\As'_\kappa }
\def \as {{ \bf as          }}

\def \ask { \as_\kappa      }
\def \ass { \as_\sigma      }
\def \asks{ \as_{\k,\s}     }
\def \T  {{ \bf T           }}
\def \Tk  { \T_\kappa       }
\def \Ts  { \T_\sigma       }
\def \Tm  { \T_m            }
\def \Tmp { \T_{m'}         }
\def \t  {{ \bf t           }}
\def \tk  { \t_\k           }
\def \ts  { \t_\s           }
\def \P  {{ \cal P          }}
\def \D  {{ \cal D          }}
\def \S  { { \cal S }       }
\def \k   { \kappa          }
\def \d   { \delta          }
\def \l   { \lambda         }
\def \L   { \Lambda         }
\def \s   { \sigma          }
\def \o   { \omega          }
\def \Smax { S_{\rm max} }
\def \.{\raise0.3ex\hbox{$\scriptscriptstyle\circ$}}
\def \asy #1#2{\ \displaystyle\mathop\simeq_
                  {{\displaystyle\mathstrut}#1\to#2}\ }
\def\emtyset{{\mathchoice{\not{\kern-0.15em\circ}}{\not{\kern-0.15em\circ}}{\not{\kern0.1em\circ}}{\not{\kern0.09em\circ}}}}
\def\etyset{{\mathchoice{\not{\kern-0.2em\lower0.035em\hbox{\xivpt$\textstyle\circ$}}}%
                        {\not{\kern-0.2em\lower0.035em\hbox{\xivpt$\textstyle\circ$}}}%
                        {\not{\kern0.07em\lower0.02em\hbox{\xivpt$\scriptstyle\circ$}}}%
                        {\not{\kern0.095em\hbox{\xivpt$\scriptscriptstyle\circ$}}}}}
\let\emptyset\etyset
\def \sumx #1#2{\mathop{\displaystyle\sum_{#1}}_{\hbox{\scriptsize#2}}}
\def \abs    #1{\left|#1\right|}
\def \ang    #1{\langle#1\rangle}
\def \absang #1{\abs{\ang{#1}}}

\makeatother

\begin{document}

\def\Cref#1{#1\typeout{! External reference `#1'}%
\typeout{l.\the\inputlineno}}
\newlabel{I/ss8.3}{{19.3}{98}}
\newlabel{I/ssi1.2}{{7.8}{35}}
\newlabel{I/ss4.5}{{10.9}{54}}
\newlabel{I/ss7.4}{{18.4}{96}}
\newlabel{I/ss3.6}{{17.4}{90}}
\newlabel{I/ss3.8}{{17.5}{92}}
\newlabel{I/s4}{{13}{68}}
\newlabel{I/ss4.2}{{10.7}{52}}
\newlabel{I/ss1.7}{{15.7}{82}}
\newlabel{I/ssi4.2}{{7.2}{31}}
\newlabel{I/si3}{{6}{29}}
\newlabel{I/ssi6.2}{{11.2}{57}}
\newlabel{I/ss9.1}{{20.1}{99}}
\newlabel{As.Uniqueness}{{15.4}{81}}
\newlabel{e2.5}{{7.10}{34}}
\newlabel{e2.6}{{14.1}{73}}
\newlabel{e2.7}{{14.2}{73}}
\newlabel{e2.8}{{14.5}{74}}
\newlabel{e2.11}{{14.7}{74}}
\newlabel{e2.17}{{14.13}{75}}
\newlabel{e2.18}{{14.14}{75}}
\newlabel{ss2.1}{{7.5}{33}}
\newlabel{ss2.3}{{7.6}{34}}

\def\1{\cite{Volume.I}}
\def\2{\cite{Volume.I}}

\thispagestyle{empty}
\hbox to\hsize{\hfill \tensl PREPRINT INR-809/903}
\hbox to\hsize{\hfill \tensl APRIL 1993}
\vbox to2.8cm{\vfill}         
\begin{center}
{\large\bf
     TECHNIQUES OF DISTRIBUTIONS\\
[1mm]
     IN PERTURBATIVE QUANTUM FIELD THEORY.\\
[2mm]
    {(II) \normalsize\bf Applications to Theory of Multiloop Diagrams.}\\
[6mm]}
{\bf A.~N.~Kuznetsov}
{\bf and F.~V.~Tkachov}
\\[2mm]
Institute for Nuclear Research of the Russian Academy of Sciences,\\
Moscow 117\,312, Russia\\
[15mm]
\end{center}

{
\centerline{ABSTRACT}
\vskip2mm
\noindent
The results of the mathematical theory of asymptotic operation
developed in\1 are applied to problems of immediate physical interest.
First, the problem of \tUV\ renormalization
is analyzed from the viewpoint of asymptotic behaviour of integrands
in momentum representation.
A new prescription for UV renormalization in momentum space representation
is presented (generalized minimal subtraction scheme);
it ensures UV convergence of renormalized diagrams by construction,
makes no use of special (e.g. dimensional) regularizations,
and comprizes massless renormalization schemes (including the MS scheme).
Then we present formal regularization-independent
proofs of general formulae for
Euclidean asymptotic expansions of renormalized Feynman diagrams
(inlcuding short-distance OPE, heavy  mass expansions and mixed asymptotic
regimes etc.) derived earlier in the context
of dimensional regularization\cite{I},\cite{Inew},\cite{II}.
This result, together with the new variant of \tUV\ renormalization,
demonstrates the power of the new
techniques based on a systematic use of the theory of distributions and
establishes the method of \tAs-operation as a comprehensive
full-fledged---and inherently more powerful---alternative
to the BPHZ approach.
}

\setcounter{footnote}{0}
\newpage

\section{Introduction}

In the preceding paper\1
we undertook a regularization-independent formalization of the
heuristic reasoning behind a series of
publications\cite{fvt:q82},\cite{r*},\cite{fvt:83},\cite{algorithm},\cite{I},\cite{Inew},\cite{II}
in which efficient methods of perturbative calculations were found
(for references to various 2-, 3-, 4- and 5-loop calculations performed using
that techniques see\cite{paradigm}).
The new techniques for studying multiloop Feynman diagrams
is based on a systematic use of the ideas of the distribution theory,
and the key notion is that of asymptotic expansion in the sense of
distributions\cite{fvt:q82},\cite{I},\cite{Tk-V:2}.
A very general context in which to construct
such expansions is established
by the {\em extension principle\/}\cite{fvt:q82},\cite{Tk-V:2}---an
abstract functional-analytic proposition analogous to the classic
Hahn-Banach theorem.
A specific realization of the recipe implied
by the extension principle---and the key instrument
of our techniques---is the so-called
asymptotic operation (\tAs-operation)\cite{I},\cite{Tk-V:2}.
The Euclidean version of
\tAs-operation constructed in\cite{I},\cite{Tk-V:2} is defined
on a class of products of singular functions
comprising integrands of Euclidean multiloop Feynman diagrams.
The \tAs-operation returns their expansions in powers and logarithms
of a small parameter (e.g., a mass) in the sense of distributions.

In the first paper\1 an analytic technique was developed
for describing singularities of distributions,
as well as a combinatorial formalism
({\em universum of graphs\/}) to work with hierarchies of graphs and their
subgraphs. Such a formalism makes it easy to utilize inherent recursive
structures in problems involving multiloop diagrams. As a warm-up exercise,
a very  compact proof of a (localized) version of the familiar
Bogoliubov-Parasiuk theorem in coordinate representation was presented
with a  purpose  of  illustrating  in detail the typical ways of
reasoning within the new techniques.

It was constructively proved that the asymptotic expansions
in  powers and logarithms of the small parameter
exist for a large class of  products  of singular functions that
includes  integrands  of  Euclidean   Feynman diagrams in   momentum
representation. The simplest example of such an expansion is
with  respect to a mass in a product of propagators, while the
\tAs-operation applied to such a product yields asymptotic expansions
in powers and logs of the small  mass, with coefficients given by
compact explicit expressions suitable for studying expansion problems
within the framework of applied Quantum Field Theory.

In the present paper we apply that general techniques to
studying expansions of integrated Euclidean Feynman diagrams
in momentum representation.

Similarly to the first part of the review\1, our purpose here
is not to construct a complete perturbation theory
or prove a short-distance expansion for a particular model in full detail
but rather to demonstrate the general techniques of our formalism.
Thus we do not attempt to write out, say, criteria for UV convergence
of non-renormalized diagrams; such results are well-known and present
little interest {\it per se}, and can be obtained in a straightforward
manner from our general formulae whenever needed in a specific situation.
Instead, we focus on how the techniques of the \tAs-operation
allows one to deal with the essential analytical aspects of such problems
and to avoid combinatorial complexities of the usual approaches.

First of all, when performing integrations over infinite momentum space
one encounters the well-known phenomenon of \tUV\  divergences that
are due---in one of the possible interpretations---to
slow decrease of Feynman integrands at infinite integration momenta.
However, using the results of\1
one can explicitly extract those and only
those  terms  in the asymptotic expansion
of the integrand at infinite integration momenta,
that are responsible for \tUV\ divergences.
It  turns out that such terms follow (with the opposite sign)
the pattern of the  terms to be introduced by the standard $R$-operation.
Therefore, a direct subtraction of such terms from the integrand (with
some natural precautions) results in a correct \tUV\  renormalization.

The \tUV\  finiteness of our construction is ensured {\em by definition\/}
and what has to be proved is its equivalence to the standard formulation of
the  $R$-operation.\footnote{
Which, strictly speaking,
is superfluous because one could prove correctness of
the new representation directly, without reducing it to the standard
construction. Such a proof would require
results like existence of short-distance OPE (see e.g.\cite{Brandt});
cf. also\cite{Lowen})
which is also considered in the present paper.
}
The class of subtraction schemes  that  naturally
corresponds  to  the  new  definition of UV renormalization
(the so-called  {\em generalized minimal subtraction schemes\/},
or GMS schemes) comprises  all massless schemes
(including the MS scheme\cite{tHooft:MS})
characterized by an extremely important property of polynomiality
of the renormalization group functions in masses\cite{Collins:Ren}.

A  definition  of  the  ultraviolet $R$-operation  as  a   procedure   of
subtraction of asymptotics from the momentum  representation  integrands
was first given by D.~Slavnov et al.\cite{Slavnov:GMS}.
However, in\cite{Slavnov:GMS}  such
subtractions  are performed recursively with respect  to  loops,  i.e.\
with  respect  to  each integration momentum in turn, while in our
approach asymptotics with  respect to the entire collection of loop
momenta are subtracted. This and the use  of the \tAs-operation for
products of singular functions\1 in our approach allow one
to easily  exhibit the  pattern  of subtractions that  is
characteristic  of  the  standard  definition  of the $R$-operation.

Our definition of the $R$-operation was first published in a complete form
in\cite{Kuz-Tk:nikhef}. It develops an idea from\cite{I}
and emerged from  a  study  of  diagrammatic
interpretation  of  the non-trivial  terms generated
by the \tAs-operation, not without an influence of\cite{Slavnov:GMS}.

An extremely important observation is that  the
coefficients of the \tAs-operation constructed in\1 turn out to be
exactly renormalized Feynman diagrams corresponding  to
subgraphs of the initial diagram.
This fact has a dramatic technical impact on
the problem of asymptotic expansions of Feynman diagrams,
because it allows one to restore global OPE from expansions
of individual diagrams in a very simple and straightforward fashion
without any of the complexities of the combinatorial techniques
of the BPHZ method.\footnote{
It should be noted that the combinatorial constructions
that naturally emerged within the framework of
the theory of \tAs-operation\cite{II} (in particular, the inverse
$R$-operation) caused an overhaul of how the combinatorial aspects
are treated in the BPHZ theory (see\cite{Smirnov:book})---even if
the basic nature of the old approach (a complete resolution of
all recursions) does not allow one to avoid having to deal with
the rather cumbersome multiple summation formulas etc.}

The second application of the theory of \tAs-operation
that we consider addresses the problem of
Euclidean asymptotic expansions of renormalized multiloop diagrams.
We present a compact and straightforward
derivation of general Euclidean asymptotic expansions
in the form of \tAs-operation for integrated
diagrams that was first introduced in\cite{II}.
Then the combinatorial techniques developed in\cite{II}
immediately allows one to obtain expansions for perturbative Green functions
in OPE-like form.

The importance and feasibility  of  the  general  problem  of
Euclidean asymptotic expansions was realized in\cite{I},\cite{Inew},\cite{II}.
In  those papers, a compact derivation of closed general formulae
for such expansions was presented.
The derivation of\cite{I},\cite{Inew},\cite{II},
however, aimed at obtaining the results in a  shortest way
and in a form immediately useful for phenomenological  applications,
so that a heavy use was made of the dimensional regularization and the
MS scheme\cite{tHooft:MS}.\footnote{
One can ponder on the tremendous
heuristic potential  of  the  dimensional regularization  and  the  MS
scheme.  Although  the  understanding  of the analytic  aspects  of
the problem---including   existence of the representation of
\tUV\  renormalization in the GMS form---was hardly lacking
in \protect\cite{I},\cite{Inew},\cite{II},
the compact presentation given there turned out feasible
due to compactness of the formalism of dimensional regularization and its
property to nullify certain types of scaleless integrals.
}
This left an open question of to what extent the
results  of\cite{I},\cite{Inew},\cite{II} are independent of regularization.

Apart from the general interest, there are also very practical reasons
for developing a regularization independent formalism.
First of all, there are the notorious
difficulties that the dimensional regularization encounters  when
applied to models involving $\gamma_5$  or supersymmetry.
In particular, the $\gamma_5$ problem emerges if one wishes to use
the chirality representation-based formalisms that are used to facilitate
the enormously cumbersome gauge algebra
in calculations of radiative corrections in QCD
(cf.\ e.g.\ the superstring theory-inspired formalism developed in
\cite{Kosower}).

Another reason is the breakdown of dimensional regularization
in non-Euclidean asymptotic expansion problems\cite{Col-Tk:Breakdown},
so that the regularization-independent formalism seems to be the only basis
for construction of practical algorithms in the non-Euclidean case.
This effect is connected with the insistence on expansions in
``perfectly factorized" form, which is important for the following reasons:

\begingroup
\emergencystretch\hsize
It was an important realization
of\cite{fvt:83},\cite{I},\cite{Inew},\cite{II}
that a proof of any asymptotic  expansion---be  it
Wilson's OPE  or  a heavy-mass expansion or asymptotics of the quark
formfactor in the Sudakov regime---is phenomenologically
irrelevant unless  the result exhibits {\em perfect factorization\/}
of large and  small parameters.  At the technical   level   of
diagram-by-diagram  expansions,  perfect factorization   means   that
the expansions run in pure powers and logarithms of the expansion
parameter.  Such expansions possess the property of uniqueness
(cf.\  the  discussion  in\1,~\SUB\cref{15.4})
which is tremendously useful from  the  technical  point of view;
for example, one immediately obtains that the \tAs-operation
commutes with multiplications  by  polynomials (see\1  and\cite{Inew}).
Another example is that one need not
worry about properties like gauge invariance of the expansion in a
given approximation:  such properties are inherited by the expansion
termwise from the initial amplitude, provided the expansion is
``perfect" in the above sense.\footnote{
As was pointed out to us by
J.C.~Collins, such a property should be even  more important for the
problem of asymptotic expansions in Minkowskian regimes
where both gauge invariance
plays a greater role for phenomenological reasons, and the expansions one
has to deal with are considerably more complicated.
It may be said that the ``relentless pursuit of perfection" (in the above
sense) is one of the characteristic differences of
the philosophy of the \tAs-operation from the old BPHZ paradigm.
}

For the above reasons, we consider it our major task  to
clarify the issue of existence of ``perfect" expansions in
regularization-independent way.
\endgroup

It is interesting that the derivation of  OPE  and,  more
generally, Euclidean asymptotic  expansions  presented  in  this
paper---being  more formalized than
that of\cite{I},\cite{Inew},\cite{II}---leads to a final
formula which is much easier to deal with at the final stage of obtaining
expansions for Green functions in a global ``exponentiated" form.
For example, unlike\cite{II},
we don't have to study  inversion of the $R$-operation.\footnote{\ignorespaces
Although we do use it in establishing---for the purposes of
illustration---a connection of our GMS prescription to the standard MS scheme.
}

However, as was stressed in\cite{I}, the derivation  presented
there  was geared to the calculational needs of applied Quantum Field
Theory (primarily, applications to perturbative Quantum
Chromodynamics)  and,  therefore,  dealt explicitly with \tUV\
counterterms etc. From practical point of view, the
formalism of the present paper offers, at least in its current  form,
no advantages as compared with the explicit recipes
of\cite{I},\cite{Inew},\cite{II}---{\it provided} one can perform
the calculations within dimensional regularization.

From  theoretical  point of view the formalization
undertaken in the present series is more than just  an exercise  in
rigour:  there is the major unsolved problem of asymptotic expansions
in non-Euclidean regimes, and it seems to be
intrinsically intractable by the BPHZ
method.\footnote{
Because the pattern of recursions involved is so much more complicated
than in the Euclidean case
that it does not seem possible to resolve them in an explicit
manner\cite{Tkachov:Advanced}.
}
On the other hand, extension of the \tAs-operation to
non-Euclidean regimes---taking into account the accumulated
experience\cite{sterman} which only needs to be properly organized
within an adequate technical framework---seems to be a matter of near future.
We expect the experience gained in Euclidean problems to play
a crucial role in the more complicated cases\cite{Anonymous}.

One of the main points in any proof of OPE is to
study the interaction of \tUV\  renormalization and the expansion proper.
As we are going to show, within our formalism the problem reduces
to double asymptotic expansions in the sense of distributions.
Indeed, a renormalized Feynman diagram in the GMS formulation has
a form of an integral of the remainder of the \tAs-expansion of  the
integrand in the regime when all the dimensional parameters of
the diagram are much less than the implicit \tUV\ cutoff.
When one applies the second expansion with respect to some
of the diagram's masses or external momenta, there emerges,
essentially, a double \tAs-expansion. All one has to prove is that
the double expansion thus  obtained factorizes into a composition
of two commuting \tAs-expansions and that the remainder of such a double
expansion is bounded by a factorizable function of the small parameters.
This is done by a straightforward extension of the analytical
techniques of\1.

\paragraph*{The plan of the paper}
is as follows (as in\1, sections contain a preamble
where further comments on its contents and results can be found).

In section~\sect{R.UV.Motivations} we present motivations
and a  definition  of  an
operation $\cR$ which subtracts asymptotics at large loop momenta
from  the integrand  of  a multiloop diagram,  thus  ensuring \tUV\
finiteness  of the  latter.
In section~\sect{rf.Definition} the explicit expressions
for the  \tAs-operation  from\1
are used to (partially) restrict the  arbitrariness  in  the  definition
of  the operation $\cR$.  For the purposes of illustration,
section~\sect{R.UV.MS-scheme} establishes  equivalence  of $\cR$
(with  the  arbitrary constants properly fixed) and the $R$-operation
in the MS scheme.

In section~\sect{R.UV.Structure} we obtain a useful representation
for the operation $\cR$ and apply it to deriving
a ``renormalization-group transformation"
of the $\cR$-renormalized diagrams. This transformation has exactly
the form that is characteristic of the $R$-operation.

In section~\sect{Overview} the problem
of the asymptotic expansion of renormalized Feynman diagrams is
reviewed, its heuristic analysis from the point of view of \tAs-operation
is given, and a recursive expansion formula is obtained.
Section~\sect{Structure} is devoted to combinatorial analysis of
that formula.
In section~\sect{Diagrammatic} a convenient expression for expanded
renormalized diagrams is obtained in a combinatorial form similar to
\tUV\  $R$-operation,
and its exponentiation on perturbative Green functions is considered.
All analytic details are dealt with in section~\sect{Double}
where the key theorem on double asymptotic expansion is
formulated and proved.
Some less important technical results are relegated to
two appendices.

For simplicity we will assume that the external momenta of the  diagrams
to be renormalized are fixed  at  non-exceptional  values.  This
assumption, however, is inessential, and our results can  be  extended  to
the  case  of diagrams considered as distributions with respect to their
external momenta.

The  notations  used in the present paper are the same as in\1.

\section{Motivations and definition of the operation $\protect\cR$\EOH.}
\label{R.UV.Motivations}

When studying the \tUV\  behaviour of momentum space Feynman integrands  one
normally invokes the Weinberg theorem\cite{Weinberg}
which supplies sufficient
criteria for convergence of multidimensional integrals  over  infinite
regions.  That famous theorem, however, is very general and does not in
the least take  into account specific properties of Feynman diagrams. It
turns  out  possible  to effectively reduce the problem to one-dimensional
integrals  if  one  employs the techniques  of  the  \tAs-operation  for
products  of  singular  functions developed in\1.

After fixing some notations in~\subsect{ss1.1}, in~\subsect{ss1.2}
we  show  how  the \tAs-operation emerges in the
study of \tUV\   convergence  of  Feynman diagrams.
In~\subsect{ss1.3} we introduce a  new  definition  for  the procedure
of elimination of \tUV\  divergences (the  operation $\cR$) which  will
be  shown  in subsequent sections to be equivalent to the standard
$R$-operation.

\subsection{The momentum space integrand\EOH.}
\label{ss1.1}

Let $G(p_\G,\kappa_\G)$ be the momentum-space integrand of an
$l$-loop 1PI unrenormalized Feynman diagram, which depends on  the  set
of $D$-dimensional (Euclidean) integration momenta
$p_\G=(p_1,\ldots p_l)$
(the aggregate variable $p_\G$ runs over the linear space $P_\G$ of
$D\times l$ dimensions)  as  well  as  on  some  external parameters
(masses and momenta) that are  collectively  denoted  as $\kappa_\G$.
For simplicity the external momenta are also  taken  to  be  Euclidean but
this limitation is inessential (cf.~\subsect{I/ss8.3} of\1).

When studying the expression $G(p_\G, \kappa_\G)$  it  is  convenient
to  use  the formalism and notations developed in\1. In
particular, $G(p_\G, \kappa_\G)$ can  be considered as a special case of
the graph in the sense of\1
(cf.\  examples in~\subsect{I/ssi1.2} in\1).
Such special graphs (i.e.\  corresponding to integrands of
Feynman diagrams in momentum representation) can  be  conveniently  called
$F${\it-graphs\/}. To distinguish subgraphs in the sense of\1 from
the 1PI subgraphs that are used in the theory  of  the $R$-operation,
we  will  use  the  terms \IR-{\it subgraphs\/} and \UV-{\it subgraphs\/},
respectively.

The notations ${\rm IR}[G]$ and ${\rm UV}[G]$ will be used to denote
the sets of all \IR- and \UV-subgraphs  of  a  given $F$-graph  $G$.
(In  the notations  of\1,
${\rm IR}[G] = S[G]$; we  will  use  the  term
$s$-subgraph  when  discussing  general properties of \tAs-operation,
while the  term \IR-subgraph  is  used
when \UV-subgraphs enter into consideration.)

We wish to consider convergence of the integral
\be[eXXXX]
\int dp_\G\, G(p_\G, \kappa_\G) \EOE{}

\ee
at $p_\G\to\infty$.

\subsection{Reduction to \protect\tAs-expansion problem\EOH.}
\label{ss1.2}

Introduce  into  the  integrand  of \eq{eXXXX}
an   arbitrary   smooth   cutoff   function
\be
\Cut(p_\G/\Lambda)=\int^\Lambda_0 d\lambda/\lambda\, \cut(p_\G/\lambda)\EOE,

\ee
where $\Cut(p_\G) \in  {\cal D}(P_\G)$ and
$\Cut(p_\G)=1$ in  a  neighbourhood of $p_\G=0$,
and $\cut(p_\G)$ differs from zero in a spherical layer.  Then  change  the
order of integrations over $\lambda$ and $p_\G$:
\be[e1.1]
 \int dp_\G\, \Cut(p_\G/\Lambda) G(p_\G,\kappa_\G)
 = \int^\Lambda_0 d\lambda/\lambda\,
     \Bigl[
       \int dp_\G\, \cut(p_\G/\lambda) G(p_\G,\kappa_\G)
     \Bigr] \EOE.

\ee
We see that the behaviour  of  the  l.h.s.\   at
$\Lambda\to\infty$  is  determined  by  the behaviour at
$\lambda\to\infty$ of the square-bracketed expression  on  the  r.h.s.\
which, after rescaling $p_\G \rightarrow  \lambda p_\G$
and  taking  into  account  that $G(p_\G,\kappa_\G)$
has  a definite dimension $\dim  G$, takes the form:
\be[e1.2]
\lambda^{\omega^\0_\G} \int  dp_\G\, \cut(p_\G) G(p_\G, \kappa_\G/\lambda)\EOE.

\ee
Here $\omega_\G=D\times l - \dim G$ coincides with the usual
index of \tUV\  divergence.

So, we have reduced the problem of studying asymptotics of
$G(p_\G, \kappa_\G)$ at large momenta to studying the asymptotic
expansion at $\lambda\to\infty$ of  the  integrals \eq{e1.2}
with test functions $\cut(p_\G)$ which are zero in neighbourhoods of
$p_\G$=0,  in other words, to the problem of asymptotic expansion at
$\lambda\to\infty $ of the expression
$G(p_\G,\kappa_\G/\lambda)$ considered as a distribution over test
functions  from  the  space ${\cal D}(P_\G \backslash \{0\})$. Up to a
redefinition of the expansion parameter $\kappa_\G/\lambda \rightarrow
\kappa $, this is a special case of the more general problem which was
formulated and  explicitly solved in\1.
From\1 it follows
that there exists (and can be presented in an explicit form---see
eq.\cref{(20.16)} in\2) a  unique asymptotic  expansion  of
$G(p_\G,\kappa_\G/\lambda)$ in powers and logarithms of $\lambda$ in
the  sense  of  the  distribution theory:
\be[e1.3]
G(p_\G,\kappa_\G/\lambda)
\mathrel{\mathop{\simeq}_{\lambda\to\infty}}
\As'_{\kappa^\0_\G}\.G(p_\G,\kappa_\G/\lambda)
= \sum_{n\geq 0}\lambda^{d-n} \sum_i C_{n,i}(\ln \lambda)
G_{n,i}(p_\G)\EOE.

\ee
Here $C_{n,i}(\ln \kappa)$  are  polynomials  of $\ln  \kappa,
G_{n,i}(p_\G)$  are  distributions  on ${\cal D}(P_\G \backslash \{0\})$,
and $d$ is a numeric constant. (The expansion which is valid on
${\cal D}(P_\G)$ is denoted as $\As$. It is not used in the
context of the problem of \tUV\ convergence.)

Each term in the sum over $n$ in \eq{e1.3} inherits the dimensional
properties of the l.h.s., so  that  a  straightforward  power  counting
allows  one  to determine  exactly  which  terms  on  the  r.h.s.\   are
responsible  for \tUV\  divergences at $\Lambda\to\infty $ in
\eq{e1.1} (namely,  the  terms  with $\omega_\G+d-n \geq 0$).  It
should be stressed that {\it any\/} \tUV\  divergence on the l.h.s.\  of \eq{e1.1}
(even  the ones corresponding to divergences in subgraphs in  the  case
of  a  negative index for $G$ as a whole), will give rise in \eq{e1.3} to a
term  resulting  in  a divergence after integration at large $\lambda $ in
\eq{e1.1}. Subtraction of  such  terms from $G(p_\G,\kappa_\G/\lambda)$
will ensure \tUV\  convergence of \eq{e1.1}.

\subsection{Definition of the operation $\cR$\EOH.}
\label{ss1.3}

It is important to note that the  term  with $\omega_\G + d - n$ = 0
which  is responsible  in \eq{e1.3}  for  the  logarithmic   divergence   at
$\lambda\to\infty$  (or, equivalently,  at $p_\G\to\infty $
in  the  integral $\int dp_\G G(p_\G,\kappa_\G)$),  also
has   a logarithmic singularity at $\lambda\to 0$ (respectively,
at $p_\G\to 0$). This singularity is localized at an isolated
point ($p_\G=0$) and can be removed (i.e.\
$\As'_{\kappa^\0_\G}\.G(p_\G,\kappa_\G)$ can be extended
to a distribution over ${\cal D}(P_\G)$) with the help of the operators
$\r$ that were introduced and studied in\1.
Such an extension is,  of  course,
not unique, but different extensions differ only by a distribution
localized at $p_\G=0$
(i.e.\  by a superposition of derivatives of the $\delta$-function).

Let us define the operation $\cR$ which acts  on  a  Feynman  diagram
$G$  as follows:
\be[e1.4]
\cR\.G \bydef  \lim_{\Lambda\to\infty}
\int dp_\G\, \Cut(p_\G/\Lambda)
\bigl[
   G(p_\G,\kappa_\G)-\r_f\.\As'_{\kappa^\0_\G}\.
   G(p_\G,\kappa_\G)
\bigr]\EOE,

\ee
where the subscript $f$  indicates that the subtraction
operator $\r_f$ is chosen in the natural ``factorized" form
(see below).

Formally speaking, the  definition \eq{e1.4}  subtracts  the  entire
asymptotic expansion, not only the terms that are responsible for \tUV\
divergences.  But because of the so-called {\it minimality property\/}
(see\1,~\subsect{I/ss4.5}),
each of the extra terms (the  ones  with
$\omega_\G+d-n<0$) is nullified  by  termwise integration over
$p$ for all sufficiently large $\Lambda$. This remarkable
fact allows one {\it not\/} to indicate explicitly how many terms
are retained in the \tAs-operation in \eq{e1.4}---and we will normally not do
that.

It is asserted that with some natural limitations on $\r_f$ the
operation $\cR$ is equivalent to the standard $R$-operation while the
remaining  arbitrariness in  the  choice  of $\r_f$  will
correspond  to  the  ordinary  renormalization arbitrariness of the
$R$-operation. But prior to defining $\r_f$ it is necessary to
study the structure of the \tAs-operation in more detail.

{\bf Remark}. Strictly speaking, even in simple cases---e.g.\   if
there  are massless particles  in  the  model---$G(p_\G,\kappa_\G)$
may  contain  non-integrable singularities at finite $p$. A typical
and the most important example is a pair of massless propagators with the
same  momentum  flowing  through  them  and separated by a self-energy
insertion. (Such  singularities  are  suppressed after integrations over
the loop momenta of the insertion.) In this case such singularities should
preliminarily be removed  using  e.g.\   the  operation $\R$
defined  in\1 (concerning  definition  of $\As$  on  such
expression  see~\subsect{I/ss7.4} in\1) or by introducing
non-zero masses into the  corresponding lines and putting them to zero in
the final expression \eq{e1.4}.  Whether  the final result will be
affected by such a redefinition or not, depends  on  the model and on how
the \tUV\  divergences have been subtracted.  In  the  mentioned example
with the pair of massless  propagators,  no  problems  arise  if  the
insertion is zero at its zero external momentum both before and after the
\tUV\  renormalization which means that the insertion is proportional to the
squared momentum. This normally happens  in  gauge  models  with  a  gauge
invariant renormalization. If, on the other hand, the insertion does not
possess such a property,  then  the  interactions  generate  a   non-zero
mass   for   the corresponding particle and the perturbation theory which
starts with massless particles should be considered incorrect. One can
imagine  more  complicated situations, but a general analysis of  such
complications---which is more or less straightforward, but
requires taking into account specific detailes
of the particular problem---goes  beyond  the scope of the present paper.
In what
follows we will  limit  ourselves  to the case when $G(p_\G,\kappa_\G)$
is locally integrable for all finite $p_\G$---this happens e.g.\  when
$G$ does not contain massless propagators.

\section{Definition of $\protect\r_f$ and the structure of
         $\protect\r_f\.\protect\As'_{\kappa^\0_\G}\.G$\EOH.}
\label{rf.Definition}

In this section we complete the definition of the operation $\cR$,
eq.\eq{e1.4}, by fixing $\r_f$,
and  study  the  structure  of  the  resulting  expressions
using specific properties of Feynman diagrams. The technical results of
this section form a basis for a detailed study of the properties of the
subtraction procedure $\cR$ in subsequent sections.

\subsection{Counterterms of \protect\tAs-operation
            and the operation $\cR$\EOH.}
\label{ss2.6}
\label{R.FiniteCounterterms}

Recall (\subsect{I/ss3.8} in\1) that the
\tAs-operation is constructed  using  the  special operation
$\R$ with  specially
chosen  finite  counterterms  depending  on  the expansion parameter.
For the operation $\As'_\kappa$,  one has the expression
similar to eq.(\cref{20.16}) of\1 but with $\R$ and $\r$
replaced by $\R'$ and $\r'$, with summation over $\gamma$
restricted to $\gamma\prec G$. Application of $\r_f$ is then
equivalent to removing primes on the r.h.s.:
\be[e2.22]
\r_f \.\As'_\kappa \.G
 =
\sum_{\gamma\prec G}
  \R_f \.\bigl[(\E_{f,\kappa} \.\gamma)
                    (\T_\kappa \. G \backslash \gamma)
         \bigr] \EOE,

\ee
where
\be[e2.20]
\E_{f,\kappa} \.\gamma
 =
\sum_\beta
\tilde{E}_{f,\gamma,\beta}(\kappa)\, \delta^\beta_\gamma\EOE.

\ee
The explicit expression for $\tilde{E}$ is given by~\cref{(20.5)} in\1
which, taking into account the definition of $\cR$, can be
rewritten as
\be[e2.24]
\tilde{E}_{f,\gamma,\beta}(\kappa)
\equiv
\cR \.\bigl[ {\cal P}^\beta_\gamma
                 *
                 \gamma
      \bigr]
\equiv
\cR \.\Bigl[
                 \int dp_\gamma\,
                   {\cal P}^\beta_\gamma(p_\gamma)
                   \gamma (p_\gamma, \kappa)
      \Bigr]\EOE.

\ee
It  is
sufficient  to notice that the multiplication by a polynomial commutes
with $\As'$ owing to the uniqueness of the latter
(\subsect{I/ss1.7}  in\1),  and  with $\r_f$  up  to
a variation of the finite counterterms.

\subsection{\UV-subgraphs as co-subgraphs\EOH.}
\label{ss3.1}

Now, return into the  context  of \eq{e1.4}.  Then $G$  is  an $F$-graph
and eqs.\eq{e2.22}--\eq{e2.24} remain valid with $\kappa$ replaced by
$\kappa_\G$.

We begin with a combinatorial observation. The summation in \eq{e2.22}
runs over all $s$-subgraphs $\gamma$ (which in the present case are called
\IR-subgraphs) of the $F$-graph $G$, $\gamma \ne G$.
An \IR-subgraph $\gamma$ is an arbitrary set  of  lines
and vertices of $G$ which satisfies the completeness condition
(see~\subsect{I/ssi4.2}  in\2).
In the present case this means the
following: ({\it i\/}) when all the external momenta are nullified as well as
all the momenta flowing through the lines of $\gamma$, no other line of
$G$ will have its momentum nullified owing to the  momentum conservation
at vertices; ({\it ii\/}) $\gamma$ contains all those and only  those
vertices of $G$ (irrespective of whether or not  there  are  non-zero
external  momenta entering into them) whose all incident lines belong to
$\gamma$  (see the examples of \IR-subgraphs in Fig.1 and Fig.3, where such
vertices are the fat ones).

Consider the complement of $\gamma$ in $G$, denoted as $G \backslash\gamma$
referred to as co-subgraph in~\SUB\cref{7.5} of\1.
The  graphical  image for $G \backslash\gamma$ is
obtained by deleting the lines and vertices belonging  to $\gamma$  from
the diagram $G$. It is not difficult to see (cf.\  the example in
Fig.3) that the connected  components  of $G \backslash\gamma$  are
precisely \UV-subgraphs  of  $G$. (By  an \UV-subgraph we mean a
subset of vertices of $G$ together with some of the lines connecting them,
and the \UV-subgraph must be 1PI and  possess  at  least  one loop. The
standard definition of the $R$-operation
in the MS scheme\cite{Collins:Ren} uses such subgraphs.
The \UV-subgraphs are defined to be non-intersecting
if  they have no common vertices.) Conversely, the complement of any set
of  pairwise non-intersecting \UV-subgraphs $\{\xi_i\}$, $\xi_i\ne G$,
is  a  correct \IR-subgraph.  Thus, there is a one-to-one correspondence
between \IR-subgraphs $\gamma$ of the $F$-graph $G$ and sets of its
non-intersecting proper \UV-subgraphs, which is  expressed  by the
following relation:
\be[e3.1]
G\backslash\gamma = \prod_i \xi_i \EOE.

\ee
Hence, there are natural factorization structure in
graphs that emerge in the context of perturbative
quantum field theory.
This simple fact is a key to establishing equivalence of the operation
\eq{e1.4} with $\r_f$ defined in \eq{e2.22} to the standard
$R$-operation.

Denote the set of the loop momenta of the \UV-subgraph $\xi_i$  as
$p_{\xi_i}$.  Then the variable $p_\G$ is split as follows (cf.\ Fig.3):
\be[e3.2]
p_\G = (p_\gamma, p_{\xi_1}, \ldots, p_{\xi_i}, \ldots)\EOE,

\ee
where $p_\gamma$ are the proper variables of the \IR-subgraph $\gamma$
(see~\SUB\cref{6.1} in\2). Then
\be[e3.3]
G(p_\G,\kappa_\G) \equiv  \gamma (p_\gamma,\kappa_\G) \prod_i
\xi_i(p_{\xi_i},p_\gamma,\kappa_\G)\EOE.

\ee
It remains to notice that if $\kappa_{\xi_i}$ is the set of all external
parameters of the \UV-subgraph $\xi_i$ (i.e.\  masses in its lines, and
the momenta that  are  external to $\xi_i$ irrespective of whether they
are external or internal for $G\supset\xi_i$), then
\be
\kappa_{\xi_i} = (p_\gamma, \kappa_\G)\EOE.

\ee

It is also convenient to introduce the notation
\be[e3.4]
p_\Gg
\bydef
(p_{\xi_1}, \ldots, p_{\xi_i}, \ldots),
\qquad
\kappa_\Gg
\bydef
(p_\gamma, \kappa_\G)\EOE,

\ee
so that
\be[e3.5]
G(p_\G, \kappa_\G)
\equiv
\gamma(p_\gamma, \kappa_\G)
\times
G\backslash\gamma
(p_\Gg, \kappa_\Gg)\EOE.

\ee

\subsection{Construction of $\protect\r_f\.\As\.G$\EOH.}

Let us specify the constructions of~\subsect{R.FiniteCounterterms}
to the present case. First consider the operation $\R$.

Consider the collection of all 1PI Feynman diagrams.  If $G$  belongs
to this collection then we denote its momentum  space  integrand  as
$G(p_\G,\kappa_\G)$, where $p_\G$ is the set of its loop  momenta  and
$\kappa_\G$  denotes  the  set  of  its external parameters, i.e.\
masses and external momenta. $G(p_\G,\kappa_\G)$ is an $F$-graph in the
sense of~\subsect{ss1.1}. All the \UV-subgraphs of $G$  also
belong  to  the same collection.

Consider expressions of the form
$\bigl[{\cal D}^\alpha_{\kappa^\0_\G} G(p_\G,\kappa_\G)\bigr]_{\kappa^\0_\G=0}$
where ${\cal D}$ are  partial
derivatives and $\alpha$ is a multiindex. Such expressions can be
transformed  into well-defined distributions over $p_\G$  using  special
operations $\R$  that  were introduced in\1.

Assume that for each $G$ from our collection such
$\R=\R_{(G)}$ has somehow  been fixed (we are using  the
same  conventions  as  in\1
concerning  the subscripts  in
brackets  denoting (sub)graphs---cf.~\SUB\cref{4.3} in\1).   In
particular,  for  each \UV-subgraph
$\xi(p_\xi, \kappa_\xi) = \xi(p_\xi, p_\Gx, \kappa_\G)$
from $G$   the expressions
$\R_{(\xi)}\.{\cal D}^\alpha_{\kappa^\0_\mxi} \.\xi
\bigr|^\0_{\kappa^\0_\mxi=0}$
are well-defined distributions over $p_\xi$. We
need not---and {\em will not\/}---assume that there is any  connection
between $\R_{(G)}$  and $\R_{(\xi)}$, $\xi \in {\rm UV}[G]$,
except for the special case when $G$ is factorizable
and $\xi$ is one of its factors
(cf.\ the definition $\R$ on factorizable
graphs in~\SUB\cref{11.2} of\1).
In the latter case, the subgraph $\xi$ is automatically an \IR-subgraph
(cf.~\SUB\cref{7.6} of\1).

{\emergencystretch0.5\hsize
If $\T_{\kappa^\0_\G}$ is the Taylor expansion in $\kappa_\G$ then the
expression $\R\.\T_{\kappa^\0_\G} \.G(p_\G,\kappa_\G)$
implies termwise application of $\R=\R_{(G)}$  which  by
definition  commutes  with powers of $\kappa_\G$.
}

In order to use the results of~\subsect{R.FiniteCounterterms},
an operation $\R=\R_{(G \backslash \gamma)}$
should be defined for $G\backslash\gamma$ for each
\IR-subgraph $\gamma$.  We  may  assume
that $\R$  always satisfies the factorization condition of~\subsect{I/ssi6.2} in\2,
so that  if $G\backslash\gamma$ is
factorized into several \UV-subgraphs as in \eq{e3.5} then
\be[e3.6]
\R_{(G\backslash\gamma)}
\.
 \T_{\kappa^\0_\Gg}
 \.
   G\backslash\gamma (p_\Gg, \kappa_\Gg)
=
\prod_i
  \R_{(\xi_i)}
  \.
  \T_{\kappa_i}
  \.
  \xi_i(p_{\xi_i}, \kappa_{\xi_i})\EOE,

\ee
where $\T_{\kappa_i}\equiv\T_{\kappa_{\mxi^\0_i}}$ and we have used the
fact that the Taylor expansion also factorizes.

Now, instead of \eq{e2.8} of\1 one has:
\be[e3.7]
\R_f \.
           \bigl[ \delta^\alpha_\gamma
             \T_{\kappa^\0_\G}
             \.
             G\backslash\gamma
           \bigr]
\bydef
\sum_{\alpha=\beta +\sum_i\beta_i}
  \delta^\beta_\gamma
  \prod_i (\beta_i!)^{-1}
  \R_{(\xi_i)}
  \.
  \bigl(
    {\cal D}^{\beta_i}_\gamma
    \T_{\kappa^\0_\G} \.\xi_i
  \bigr)^\0_{p_\gamma=0} \EOE,

\ee
and instead of \eq{e2.22}:
\be[e3.8]
\r_f \.\As'_{\kappa^\0_\G} \.G
\bydef
\sum_{\gamma\prec G}
  \R_f \.\Bigl[ (\E_{\kappa^\0_\G} \.\gamma)
                     \prod_i \T_{\kappa^\0_\G} \.\xi_i
         \Bigr] \EOE,

\ee
where $\E_{\kappa^\0_\G}$ is given by \eq{e2.20} and \eq{e2.24}
with $\kappa\rightarrow\kappa_\G$.

It should be remembered that \eq{e3.8} is a definition of  the  operator
$\r_f$ which removes the non-integrable singularity at $p_\G=0$
of $\As'_{\kappa^\0_\G}\.G$.  The  latter expression is  itself  defined
uniquely,  i.e.\  it is  independent  of  how  the intermediate
renormalization is performed, which means that it is independent of
$\R'_{f(G)}$.

The above definition for $\r_f$ completes the definition of the
operation $\cR$ in \eq{e1.4}.

\subsection{``Factoring out'' $\delta$-functions\EOH.}

Let us derive, using the above results, a useful representation for  the
expression \eq{e3.8}. The point is that the factors
$\E_{\kappa^\0_\G}\.\gamma$ contain derivatives of $\delta$-functions
$\delta^\beta_\gamma(p_\gamma)$
(cf.\ \eq{e2.20}), but the factors $\xi_i$ in \eq{e3.8} may depend
on the arguments of such $\delta$-functions.
Integration of the $\delta$-functions will cause $\xi_i$ to be
differentiated in $p_\gamma$.
Therefore, where the factors $\xi_i$ in \eq{e3.8}
are differentiated only in $\kappa_G$,
they will be---after integrating out
the $\delta$-functions---differentiated in $\kappa_\G$ and
some of the components of $p_\gamma$. The following calculation
makes this effect explicit.

Taking into  account \eq{e2.20}  and \eq{e2.24},  one  obtains  the
following expression for the r.h.s.\  of \eq{e3.8}:
\be[e3.9]
\sum_{\gamma \prec G}
\sum_{\alpha}
\sum_{\alpha=\beta + \Sigma_i \beta_i}
  \cR \bigl[
        {\cal P}^\alpha_\gamma \ast \gamma
      \bigr]
  \delta^\beta_\gamma
  \prod_i
    \R_f \.\bigl(
             {\cal D}^{\beta_i}_\gamma
             \T_{\kappa^\0_\G}
             \.
             \xi_i
           \bigr)_{p^\0_\gam=0}
\EOE.

\ee
Note that the proper variables $p_\gamma$ of the \IR-subgraph $\gamma$
are divided  into  two groups of components (cf.\  Fig.3):
$p_\gamma = (p^{\rm int}_\gamma, p^{\rm ext}_\gamma)$,
where $p^{\rm int}_\gamma$ that  are
the loop momenta corresponding to the loops formed by the lines of
$\gamma$,  and $p^{\rm ext}_\gamma$ that are external for the \UV-subgraphs
$\xi_i$.
(Note that, $p^{\rm ext}_\gamma$ are  the  momenta  that  flow through the
``loose'' external lines of $\gamma$; cf.\ Fig.3.)
Furthermore, from among the components
$p^{\rm ext}_\gamma$, one can select the momenta $p^{\rm ext}_i$ that are
external with respect to $\xi_i$,  so  that  the differentiations in the
expression under $\R_f$ can be performed with respect  to
$p^{\rm ext}_i$.

The following factorization holds:
\be[e3.10]
{\cal P}^\alpha_\gamma
\equiv
p^\alpha_\gamma
=
p^\beta_\gamma \times \prod_i [p_i^{\rm ext}]^{\beta_i}\EOE.

\ee
It is easy to see that for any $\varphi(p)$ the expression of the form
$\sum_{\beta} p^{\beta} ( {\cal D}^{\beta} \varphi(p) )_{p=0}$
is simply the Taylor expansion in $p$.

Therefore, in \eq{e3.9} with \eq{e3.10}  there  will emerge,
instead  of  the  Taylor expansion of the \UV-subgraphs
$\xi_i$ in $\kappa_\G$, i.e.\ in the masses and momenta that are
external for $G$, the Taylor expansion in $\kappa_{\xi_i}$, i.e.\
the masses and momenta that are external for $\xi_i$.
It is important that the expansion in $\kappa_{\xi_i}$ emerges
irrespective of whether or not a momentum  that  is external for $\xi_i$ is
also external for the entire $G$.

Finally, using \eq{e3.9}--\eq{e3.10} we can rewrite \eq{e3.8} as follows:
\be[e3.12]
\herring{
\r_f \.\As'_{\kappa^\0_\G} \.G
\==
\R \.\T_{\kappa^\0_\G} \.G
+
\sum_{\emptyset\prec\gamma\prec G}
  \delta_\gamma
  \cR \.
  \biggl[
    \Bigl(
      \prod_i \W_i
    \Bigr)
    \ast
    \gamma
  \biggr]
\\
\> \quad +
\sum_{\emptyset\prec\gamma\prec G}
\sum_{\beta>0}
\delta^\beta_\gamma
 \cR \.
 \biggl[
   \Bigl(
     p^\beta_\gamma \prod_i \W_i
   \Bigr)
   \ast
   \gamma
 \biggr],
}
\EOE{}

\ee
where:
\be[e3.13]
\W_i(p_{\xi_i},\kappa_{\xi_i})
\bydef
\R_{(\xi_i)}\.
\T_{\kappa_i}\.
\xi_i (p_{\xi_i}, \kappa_{\xi_i})\EOE,

\ee
and the terms with $\gamma=\emptyset$ and $\beta=0$
in \eq{e3.12} have been separated from the rest of the sum.

\subsection{Remarks\EOH.}
\label{ss3.4}

To  conclude  this  section,  a  few  comments  concerning \eq{e3.12}
are necessary.
\textindent{({\it i\/})} The terms in the last sum
on the r.h.s. of \eq{e3.12} are inessential in the sense that
they can be got rid of in $\cR\.G$ (see below~\subsect{ss5.1}).
\textindent{({\it ii\/})} Recall (see~\subsect{ss3.1}) that one could replace
the summation  over \IR-subgraphs $\gamma$ by a summation over  all
sets  of  pairwise non-intersecting proper \UV-subgraphs $\{\xi_i\}$.
Such a pattern  of summation  is  typical  of  the standard
representations of the $R$-operation.
\textindent{({\it iii\/})} Diagrammatically, the expressions
\be[e3.14]
\cR \.\biggl[
		 \Bigl(
                   \prod_i \W_i
		 \Bigr)
                 \ast
                 \gamma
               \biggr]\EOE{}

\ee
are obtained by contracting the \UV-subgraphs $\xi_i$ in $G$  to
points  and  using $\W_i(\kappa_{\xi_i},p_{\xi_i})$ for the
corresponding vertex factors.
The aggregates $\W_i (\kappa_{\xi_i}, p_{\xi_i})$
are ``infinite-order" polynomials of
$\kappa_{\xi_i}=(\kappa_\G, p^{\rm ext}_i)$, the external  parameters of
$\xi_i$. Recall that the operation $*_\gamma$ involves only
integrations  over $p_\gamma$  of which each $p^{\rm ext}_i$ is a part,
and does not affect the  dependence  on $p_{\xi_i}$,  the loop momenta
of $\xi_i$, with respect to which $\W_i$ are well-defined
distributions.  Such a replacement  of \UV-subgraphs  by  polynomials  of
momenta  that  are external to the subgraphs is again typical of the
standard $R$-operation,  but in the present case the result is
``renormalized" using $\cR$.
\textindent{({\it iv\/})} Note  that  when  we  will  substitute \eq{e3.12}  into
\eq{e1.4},   the integration over $p_\gamma$ as a part of the
integration over $p_\G$  in \eq{e1.4}  will  be killed
by the $\delta$-functions $\delta^\beta_\gamma$ in \eq{e3.12}
and replaced by the
integration  over $p_\gamma$ implied by the operation $*$.
\textindent{({\it v\/})} It should be  pointed  out  that  the  coefficients
$\tilde{E}_{f,\gamma,\beta}(\kappa)$   can  be interpreted as the renormalized
graph $\gamma$ in which the connected components  of
$G\backslash\gamma$
are shrunk to points and  some  polynomials  are  inserted  instead
(cf.\ \eq{e2.24} and \eq{e3.10}).

So, we have  fully  defined  the  operation $\cR$  and  presented
explicit expressions for it, which can serve, in particular, as a starting
point  for comparison of $\cR$ with the $R$-operation in the MS
scheme.

\section{The operation $\protect\cR$ and the $R$-operation
         in the MS scheme\EOH.}
\label{R.UV.MS-scheme}

In this section we will show that the class of  renormalization  schemes
defined via the operation $\cR$ contains the $R$-operation in the
MS scheme.  First in~\subsect{ss4.1} we will formally introduce
the dimensional regularization into the definition of $\cR$
\eq{e1.4} and integrate out the $\delta $-functions in \eq{e3.12}.
Then in~\subsect{ss4.2} we will study the structure of the
``counterterms"  introduced by  the  operators $\M$  in \eq{e3.12}.  In
particular,  specific   nullification properties of the dimensional
regularization will be  used  to  show  that  a suitable choice of the
operations $\R$ results in the ``counterterms"  containing pure
poles in $D-4$. Lastly, in~\subsect{ss4.3} we show that \eq{e3.12}
is exactly  the representation of the $R$-operation in terms of the
``subtraction operator"  for the inverted $R$-operation, and then invoke
the results of\cite{II}
where $R^{-1}$ was studied and conclude
that  with  the  above  choice  for $\R$, $\cR$  is  the
$R$-operation in the MS scheme.

\subsection{Operation $\cR$ in dimensional regularization\EOH.}
\label{ss4.1}

Let us introduce the dimensional regularization into \eq{e1.4} as follows:
\be[e4.1]
\cR \.G
\bydef
\lim_{\epsilon\to 0}
\int dp^{[\epsilon]}_\G \,
\bigl[
   G(p_\G,\kappa_\G) -
   \r_f\.\As'_{\kappa^\0_\G} \.
   G(p_\G,\kappa_\G)
\bigr]\EOE,

\ee
where $\epsilon$ is the deviation of the complex parameter of the
space-time dimension $D$ from the canonical  integer  value,  while  all
quantities---integration measures, $\delta$-functions etc.---are
considered to be $D$-dimensional.\footnote{\ignorespaces
We have  in fact dimensionally
regularized the convergent expression on the r.h.s.\   under the limit
$\Lambda\to\infty$, commuted the limits with respect to
$\epsilon$  and $\Lambda$  and  let $\Lambda\to\infty$ before
taking off the dimensional regularization.  Justification  of  such  a
procedure is beyond the scope of the present paper.
Anyhow, there are all indications that such manipulations with
convergent expressions cannot give rise to problems.
Concerning the definition of dimensional regularization directly
for integrals over momentum space see\cite{Blecher:DR}.
Note that a direct formalization of the original definition
of\cite{tHooft:DR} is rather straightforward, and will
hopefully be considered in a future publication.
}

Since the coefficients \eq{e2.24} are finite by construction  and  enter
as coefficients  multiplying  well-defined  distributions,  they  can
also   be replaced by  regularized  versions.  This  means,  in
particular,  that  the ``principal value'' integration implied by  the
operation  $*$  in \eq{e2.24}  and, consequently,  in \eq{e3.12},  is
replaced   by   a   dimensionally-regularized integration.

Owing to \eq{e3.2} the integration measure factorizes:
\be[e4.2]
dp^{[\epsilon]}_\G
=
dp^{[\epsilon]}_\gamma
\prod_i dp^{[\epsilon]}_{\xi_i}\EOE.

\ee
Then, substituting \eq{e3.12} into \eq{e1.4}, one sees that the terms
with $\beta \ne 0$  are killed and the result can be represented as
follows:
\be[e4.3]
\cR \.G^\epsilon
=
G^\epsilon
-
\sum_{\{\xi_i\}}
\cR \.\Bigl(
                  \prod_i \M^\epsilon_{\xi_i} \.G^\epsilon
               \Bigr)\EOE.

\ee
where summation runs over all sets $\{\xi_i\}$ of pairwise
non-intersecting  proper \UV-subgraphs of $G$ including the one
consisting of only $G$  itself.  The  other notations used are as follows:
\be[e4.4]
G^\epsilon \bydef  \int dp^{[\epsilon]}_\G G(p_\G,\kappa_\G)\EOE;

\ee
the action of $\M^\epsilon_\xi$, where $\xi$ is a proper
\UV-subgraph in $G$,  on $G$  consists  in replacing $\xi$ in $G$ by a
vertex to which there corresponds the following factor:
\be[e4.5]
\int dp^{[\epsilon]}_\xi \,
\R_{(\xi)} \.\T_{\kappa^\0_\mxi} \.\xi(p_\xi, \kappa_\xi) \EOE;

\ee
the action of $\cR$ on the diagrams with \UV-subgraphs thus shrunk
is  defined  in the same way as in the case  of  the  entire $G$,  and  we
have  adopted  the convention that $\cR\.1 \equiv 1$
and $\cR$ is insensitive to the numerical coefficients in the vertex
factors (see the discussion of the structure of \eq{e4.5} in the  next
subsection) so that
$\cR\.\M^\epsilon_\G\.G = \M^\epsilon_\G\.G$.

\subsection{Analytic structure of counterterms\EOH.}
\label{ss4.2}

Consider the analytic structure of \eq{e4.5}.

The operation $\T_{\kappa^\0_\mxi}$ Taylor-expands the integrand of the
\UV-subgraph $\xi$  in $\kappa_\xi$, i.e.\  in all masses and momenta
that are external with  respect  to $\xi$---i.e.\  in all dimensional
parameters  on  which $\xi$  depends.  The  coefficients multiplying the
powers of masses and external momenta are  functions  of  the loop momenta
$p_\xi$ only, and as is well-known, the integration over  all $p_\xi$
in infinite bounds within dimensional regularization nullifies  such
functions.  However, in our case the operation $\R$ adds to
$\T_{\kappa^\0_\mxi}\.\xi(p_\xi,\ldots)$ some new terms  that are
obtained by replacing some groups of factors by counterterms proportional
to $\delta$-functions so as to make the resulting expression locally
integrable  at all finite $p_\xi$. It should be stressed  that
$\R$  does  not  affect---and  is insensitive to---the
powers of the masses and external momenta  generated  by
$\T_{\kappa^\0_\mxi}$, so that the coefficients of such  counterterms  can
be  chosen  to  be independent of $\kappa_\xi$. One can see  that  such
coefficients  can  be  chosen  to contain only pure singularities with
respect to $\epsilon$ (which  are  well-known  to consist of poles in
$\epsilon$; cf.\cite{tHooft:DR}) with numeric coefficients
(up to a general  factor $\mu^\epsilon$ to some integer power,
where $\mu$ is the 't~Hooft's unit of mass---which  should be
introduced to
preserve dimensionality  of  expressions  that are added to
$\T_{\kappa^\0_\xi} \.\xi (p_\xi, \kappa_\xi)$).

If a counterterm introduced by $\R$ is such that not all
integrations  over $p_\xi$ are killed by $\delta$-functions then such
term is nullified by  the  integration over all $p_\xi$ due to reasons
already discussed. So,  there  remain  only  terms that are proportional
to $\delta (p_\xi)$  and  its  derivatives.  But  the  terms  with
derivatives are also nullified by the integration over $p_\xi$ with
unit  weight.  Therefore, only terms proportional to $\delta(p_\xi)$
without derivatives survive.  One can check by power counting that the
power of the polynomials of  masses  and momenta which (polynomials)
multiply such counterterms is  exactly $\omega_\xi$,  the standard
index of \tUV\  divergence of the \UV-subgraph $\xi$.

Finally, one concludes that with the natural choice  of  the  operation
$\R$ explained above, the expression \eq{e4.5} takes the form
\be[e4.6]
\hbox{eq.\eq{e4.5}} =
\Z\left({\textstyle {1\over \epsilon}}\right)
\cdot {\cal P}(\kappa_\xi) \cdot
\mu^{\epsilon n}\EOE,

\ee
where $\Z$ is a polynomial of $1/\epsilon $ with numeric coefficients (the
most  general case of $\R_f$ only differs by finite
contributions to $\Z$), ${\cal P}$ is a  polynomial  of order
$\omega_{\xi_i}$ of masses and external momenta of $\xi_i$, while the
factor $\mu^{\epsilon n}$ (where $n$ is proportional to the number of
loops  of $\xi_i$)  is  standard  within  the dimensional regularization
and is introduced to preserve correct  dimensionality of different terms in
the sum \eq{e4.3}.

\subsection{Inverted $R$-operation\EOH.}
\label{ss4.3}

Thus, the operator $\M^\epsilon_\xi$ in \eq{e4.3} replaces the
\UV-subgraph $\xi $ by  a  vertex to  which  the  factor  of  the  form
\eq{e4.6}  corresponds.  Such  a  form  is characteristic of the
counterterms of  the $R$-operation  in  the MS  scheme.  However, the
expression \eq{e4.3}  is  not  a  correct  representation  for  the
$R$-operation because of the presence of $\cR$ which acts on the
diagram  with $\xi$'s shrunk to points. Therefore, $\M^\epsilon_\xi$
cannot be the  counterterm  operator  of  the operation $\cR$ if the
latter is to be an $R$-operation.

Nevertheless, rewrite \eq{e4.3} as follows:
\be[e4.7]
\cR \.\biggl\{
                 G^\epsilon +
                 \sum_{\{\xi_i\}}
		   \Bigl(
                      \prod_i \M^\epsilon_{\xi_i} \.G^\epsilon
		   \Bigr)
	       \biggr\}
=
G^\epsilon\EOE.

\ee
It is clear that the braced expression on the l.h.s.\  describes the
operation that is exactly the inverse for $\cR$:
\be[e4.8]
\cR^{-1}\.G^\epsilon
=
G^\epsilon +
\sum_{\{\xi_i\}}
  \Bigl(
     \prod_i \M^\epsilon_{\xi_i}\.G
  \Bigr)\EOE.

\ee
The arrangement of the operators $\M$ on the r.h.s.\ follows exactly
the same combinatorial pattern as that of the counterterm operators in
the standard representation of the $R$-operation.

The inversion of the $R$-operation was first  introduced  and  studied
in\cite{II}, where it was realized that this concept
streamlines the study of combinatorial aspects of
$R$- and \tAs-operations.
The explicit expressions for
the operators $\M^\epsilon_\xi$ in the  representation \eq{e4.8}  were
derived in\cite{II} in  terms  of  the  counterterm
operators of the $R$-operation.
Those expressions are polynomials  with  purely numerical coefficients, so
that if the counterterms of the $R$-operation  have the form \eq{e4.6}
then the counterterms of $R^{-1}$ has the same form and vice versa.  We
conclude that if $\R$ has been chosen as described after \eq{e4.5}
then $\cR$  is exactly the $R$-operation in the MS scheme, while
\eq{e4.3} is  its  representation in terms of the counterterm operator for
the inverse $R$-operation.

\section{Structure of the operation $\protect\cR$\EOH.}
\label{R.UV.Structure}

Calculations similar to those in the preceding section---integrating out
the $\delta$-functions that are explicitly present
in \eq{e3.12}---can  be  performed without reference to  dimensional
regularization.
This  is  done  below  in~\subsect{ss5.1}. The resulting
representation \eq{e5.3} for the operation $\cR$, in which the \tUV\
finiteness---in  contrast  to \eq{e1.4}---is  no  longer  obvious,
is nevertheless more convenient for other  purposes:  indeed,  it
possesses  an explicit recursion structure that can be effectively used in
proofs. Thus, in~\subsect{ss5.2}--\subsect{ss5.5}
the representation \eq{e5.3} is used to
establish the fact  that the arbitrariness in the definition of  the
operation $\cR$  is  equivalent  to introducing finite counterterms
for \UV-subgraphs in exactly the same  way  as in the case of the
standard $R$-operation. Concluding remarks concerning proofs of
correctness of the operation $\cR$ are contained in~\subsect{ss5.6}.

\subsection{Natural regularization of renormalized diagrams\EOH.}
\label{ss5.1}

Let us substitute \eq{e3.12} into the definition \eq{e1.4} and integrate
out the $\delta$-functions $\delta^\beta_\gamma$. Then the
derivatives ${\cal D}_\gamma$ (cf.\  \eq{e2.6} in\1) will switch over onto the
cut-off function $\Cut_\G \equiv \Cut$  which  is  arbitrary  except  that
$\Cut_\G \in {\cal D} (P_\G)$.
As was indicated in\2 the cutoffs can be chosen to satisfy restrictions
(\cref{8.15}), (\cref{8.16}) of\2.
Assuming such choice of the cutoffs one obtains the following expression:
\be[e5.3]
\cR\.G =
\lim_{\Lambda\to\infty}
  \biggl[
     G^{\Lambda}
     -
     \sum_{\{\xi_i\}} \cR \.
	    \Bigl(
               \prod_i \M^{\Lambda}_{\xi_i} \.G
	    \Bigr)
 \biggr]\EOE,

\ee
whose structure is completely analogous to that of \eq{e4.3} but
\be[e5.4]
G^{\Lambda} \bydef  \int dp_\G\, \Cut_\G(p_\G/\Lambda) G(p_\G,\kappa_\G)\EOE,

\ee
and the vertex factors introduced by the operators $\M^{\Lambda}$ are as
follows:
\be[e5.5]
\M^{\Lambda}_\xi \.\xi
=
\int dp_\xi \,
\Cut_\xi (p_\xi/\Lambda)
\bigl[
   \R \.\T_{\kappa^\0_\xi} \.\xi (p_\xi,\kappa_\xi)
\bigr] \EOE.

\ee
It is obviously a series in $\kappa_\xi$, i.e.\  in masses and external
momenta  of $\xi$.  However, owing to the minimality property of
$\R$ (see~\subsect{I/ss4.5} in\1)  all terms of a
sufficiently high order in this series will vanish  in  the  limit
$\Lambda\to\infty$. In fact, as can be seen from~\Sect{I/s4}  in\1, the vanishing terms  become zero for
all sufficiently large $\Lambda$ and, therefore, can  be  omitted  not
only from \eq{e5.4} but also from \eq{e5.3}. The non-vanishing terms
constitute---as  is easily seen by power counting---a polynomial in
$\kappa_\xi$ of exactly the order $\omega_\xi$, the standard index of
\tUV\  divergence.

\subsection{Finite renormalizations\EOH.}
\label{ss5.2}

Let us study the arbitrariness in the definition of $\cR$. (Recall
that  the operations $\R$ used in the construction---see e.g.\
\eq{e5.5}---are  not  defined uniquely.) We are going to demonstrate
that this  arbitrariness  has  exactly the same structure as in the case
of the standard $R$-operations.

More precisely, let the indices $a$ and $b$ mark two sets of the
operations $\R$ and, correspondingly, the two operations $\cR$.
Then,  as  we  are  going  to demonstrate shortly,
\be[e5.6]
\cR_a \.G
=
\cR_b\.G
+
\sum_{\{\xi_i\}} \cR_b \.
\Bigl(
   \prod_i \M^{ab}_{\xi_i} \.G
\Bigr) \EOE,

\ee
which has the meaning similar to \eq{e5.3} except that the operator
$\M^{ab}_\xi$ replaces the \UV-subgraph $\xi$ by a polynomial of its
masses and external momenta of order $\omega_\xi$, with finite
coefficients. (Explicit expressions for $\M$ are given in
\eq{e5.12}.) For $G$ without non-trivial \UV-subgraphs (which
corresponds to absence of \IR-subgraphs $\gamma\prec G$) eq.\eq{e5.6}
reduces to
\be[e5.7]
\cR_a \.G = \cR_b \.G + m^{ab}_\G \EOE,

\ee
which is obvious from the definitions.

In the most general case, eq.\eq{e5.6} is  proved  by  induction  using
the representation \eq{e5.3}. The reasoning can be made rather elementary
and  fully explicit with suitable notations introduced in the next
subsection.

\subsection{Hierarchy of \UV-subgraphs\EOH.}
\label{ss5.3}

The representation \eq{e5.3} relates a diagram $G$ with the diagrams
obtained from $G$ by shrinking some of the \UV-subgraphs of $G$  to
points  and  inserting some factors into the new vertices  that  are
polynomials  in  the  external parameters of the corresponding
\UV-subgraphs. Therefore, an  inductive  proof should run over the entire
set of all ``secondary" diagrams thus obtained from $G$. Let us
parameterize such secondary diagrams.

Let $\xi$ be an aggregate index that consists of two labels: the first
one, $v_\xi$, denotes an \UV-subgraph from $G$, while the second one,
$\alpha_\xi$, indicates  which polynomial of $\kappa_\xi$
(the external parameters of the \UV-subgraph $v_\xi$)  is  chosen for the
vertex obtained by shrinking $v_\xi$ to a  point ($\alpha_\xi$  runs
over  a  full linearly independent set of polynomials of $\kappa_\xi$).

Let $\{\xi\}$ be a set of such indices that corresponds to a set of
pair-wise non-intersecting \UV-subgraphs $v_\xi$, $\xi \in \{\xi \}$.
Then $G\{\xi\}$  denotes  the  diagram obtained by shrinking the
subgraphs $v_\xi$, $\xi \in \{\xi \}$, in $G$ to points  and  inserting
the $\alpha_\xi$-th polynomials instead. A summation over $\{\xi\}$
implies both a  summation over some collection of sets of pair-wise
non-intersecting \UV-subgraphs,  and the corresponding summation over all
the relevant labels $\alpha_\xi$. A  product  over $\{\xi\}$  implies
a  product  over  only  the  sets  of \UV-subgraphs,  with  the
corresponding labels $\alpha_\xi$ fixed.

It will also  be  convenient  to  denote  the \UV-subgraph $v_\xi$  as
``the \UV-subgraph $\xi$", and $\{\xi\}$, ``the set of \UV-subgraphs".

Let $\xi$ be an \UV-subgraph from $G$, and let $\{\Xi\}$ be a set of
\UV-subgraphs  such  that $\Xi \subseteq \xi$ or
$\Xi \cap \xi \ne \emptyset$ for all $\Xi \in \{\Xi\}$.
Then $\xi \{\Xi \}$ is the
\UV-subgraph in $G\{\Xi\}$  that  is obtained from $\xi$ by
shrinking to points those $\Xi$  that  lie  within $\xi$.  If  a
confusion is excluded, we will write just $\xi$ instead of $\xi\{\Xi\}$.

Finally, $\{\xi\} \succ \{\Xi\}$ means that  each $\xi$  contains  one
or  more $\Xi$'s (the corresponding $\alpha_\xi$ and $\alpha_{\Xi}$ may
be arbitrary). This relation specifies a  partial ordering in the set of
indices $\{\xi\}$ and the diagrams $G\{\xi\}$.

Using the  above  notations,  eq.\eq{e5.3}  describing  the  action  of
the operation $\cR$ on a diagram $G\{\Xi\}$ can be represented as
follows:
\be[e5.8]
\cR_a \.G\{\Xi \}
=
\lim_{\Lambda\to\infty}
\biggl\{
  G^{\Lambda}\{\Xi \}
  -
  \sum_{\{\xi\}\succ \{\Xi\}}
    \Bigl(
      \prod_\xi \M^{\Lambda a}_{\xi \{\Xi\}}
    \Bigr)
    \cR_a \.G\{\xi\}
\biggr\}\EOE,

\ee
where all $\M^{\Lambda a}_{\xi \{\Xi\}}$ are numerical  coefficients.  If
$\{\Xi\}=\emptyset$  then \eq{e5.8}  is  the renormalized graph $G$.

Eq.\eq{e5.6} that we wish to prove now takes the form:
\be[e5.9]
\cR_a \.G\{\Xi\}
=
\cR_b \.G\{\Xi\}
+
\sum_{\{\xi\} \succ \{\Xi \}}
\Bigl(
  \prod_\xi \M^{ab}_{\xi \{\Xi\}}
\Bigr)
\cR_b \.G\{\xi\}\EOE.

\ee
The expression for $\M^{ab}_{\xi\{\Xi\}}$ is given below in \eq{e5.12}.

The scenario of our proof of \eq{e5.9} is to use \eq{e5.8} in order to
reexpress $\cR_a$ on $G\{\Xi\}$ in terms of $\cR_a$ on
$G\{\xi\}$ for $\{\xi\} \succ \{\Xi\}$. Then we will use the inductive
assumption that \eq{e5.9} is valid on $G\{\xi\}$ and reexpress
$\cR_a$ via $\cR_b$. After that we will express
$\M^{\Lambda b}_{\xi \{\Xi\}}$
in terms of $\M^{\Lambda a}_{\xi \{\Xi\}}$ and
complete  the  proof.  In  the  next subsection we will study how
$\M^{\Lambda b}_{\xi \{\Xi\}}$ and $\M^{\Lambda a}_{\xi \{\Xi\}}$  are
related  and  then  prove eq.\eq{e5.9} in~\subsect{ss5.5}.

\subsection{Interrelation of counterterms in different schemes\EOH.}
\label{ss5.4}

${\M}^{\Lambda b}_{\xi \{\Xi\}}$ is represented as (cf.\  \eq{e5.5}):
\be[e5.10]
\M^{\Lambda b}_{\xi \{\Xi\}}
=
\int dp_{\xi \{\Xi\}}
\Cut_{\xi \{\Xi\}}(p_{\xi \{\Xi\}}/\Lambda)
  \bigl[
     \R_b \.\t_{\alpha_\xi} \.
     \xi \{\Xi\}(p_\xi,\kappa_\xi)
  \bigr]\EOE,

\ee
where $\t_{\alpha_\xi}$ is the differential operator with respect to
$\kappa_\xi$ that  corresponds to the $\alpha_\xi$-th polynomial in
$\kappa_\xi$ in the Taylor expansion of $\xi$  over  its  external
momenta.

Using \eq{e2.11} of \1, we can reexpress $\R_b$ in \eq{e5.10} in terms
of $\R_a$. Then we  can integrate out the resulting
$\delta$-functional  contributions  similarly  to \eq{e5.3}.
Using the properties of the cutoffs, we obtain:
\be[e5.11]
\M^{\Lambda b}_{\xi \{\Xi\}}
=
\M^{\Lambda a}_{\xi \{\Xi\}}
+
\sum_{\{\upsilon\}}
  \Z^{ba}_{\xi \{\Xi\}\{\nu\}}
  \prod_{\upsilon}
    \M^{\Lambda a}_{\upsilon \{\Xi\}}\EOE,

\ee
where the summation runs over all sets of \UV-subgraphs of $\xi\{\Xi\}$
excluding  the set consisting of $\xi$ itself but including the  empty
set (the  corresponding term in the sum is just $\Z^{ba}_{\xi\{\Xi\}}$).

We will prove the statement \eq{e5.9} together with the following
expression for the finite counterterms $\M^{ab}$:
\be[e5.12]
\M^{ab}_{\xi\{\Xi\}}= \Z^{ba}_{\xi\{\Xi\}}\EOE.

\ee
First, note that the coefficients of the transformation
$\cR_a \rightarrow \cR_b$ are related to the coefficients
of the transformation $\R_b \rightarrow \R_a$ which should be
compared  with the fact that in~\Sect{R.UV.MS-scheme} the inverse
$R$-operation emerged.

Second, eq.\eq{e5.12} explicitly relates the arbitrariness in the
definition of $\cR$ with the underlying arbitrariness in the choice
of  the intermediate operation $\R$.

\subsection{Proof of the formula \protect\eq{e5.9}\EOH.}
\label{ss5.5}

Let us prove \eq{e5.9} together with \eq{e5.12}, assuming that  they  have
been proved for all $G\{\xi\}$ with $\{\xi\} \succ \{\Xi\}$. Consider
the difference $[\cR_a - \cR_b] \.G\{\Xi\}$  and use
\eq{e5.8}:
\be[e5.13]
\herring{
\= [\cR_a - \cR_b] \.G\{\Xi\}
\\
\>\quad =
\lim_{\Lambda\to\infty}
\biggl\{
  -\sum_{\{\xi\} \succ \{\Xi\}}
  \Bigl(
     \prod_\xi \M^{\Lambda a}_{\xi \{\Xi\}}
  \Bigr)
  \cR_a \.G\{\xi\}
  +\sum_{\{\xi\} \succ \{\Xi\}}
  \Bigl(
     \prod_\xi \M^{\Lambda b}_{\xi \{\Xi\}}
  \Bigr)
  \cR_b \.G\{\xi\}
\biggr\}.
}
\EOE{}

\ee
Then express $\cR_a$ on the r.h.s.\  using the inductive assumption:
\be[e5.14]
\cR_a \.G\{\xi\}
=
\cR_b \.G\{\xi \}
+\sum_{\{\zeta\} \succ \{\xi\}}
\Bigl(
  \prod_\zeta \M^{ab}_{\zeta\{\xi\}}
\Bigr)
\cR_b \.G\{\zeta\}\EOE.

\ee
Substituting this into the r.h.s.\ of \eq{e5.13} one gets:
\be[e5.15]
\herring{
\hbox{\eq{e5.13}}
\= = \lim_{\Lambda\to\infty}
  \biggl\{
    \sum_{\{\xi\}\succ \{\Xi\}}
      \Bigl[
        \prod_\xi \M^{\Lambda b}_{\xi\{\Xi\}}
        -
        \prod_\xi \M^{\Lambda a}_{\xi\{\Xi\}}
      \Bigr]
      \cR_b \.G \{\xi\}
\\
\>\quad   -\sum_{\{\xi \}\succ \{\Xi\}}
      \Bigl(
        \prod_\xi \M^{\Lambda a}_{\xi\{\Xi\}}
      \Bigr)
      \sum_{\{\zeta\}\succ\{\xi\}}
	\Bigl(
          \prod_\zeta \M^{ab}_{\zeta\{\xi\}}
	\Bigr)
        \cR_b \.G \{\zeta\}
  \biggr\}.
}
\EOE{}

\ee
Now, the way $\M^{\Lambda b}_{\xi\{\Xi\}}$ and
$\M^{\Lambda a}_{\xi\{\Xi\}}$ enter into the r.h.s.\
indicates that there should be some
cancellations possible. Let us express $\M^{\Lambda b}_{\xi \{\Xi\}}$ in
terms of $\M^{\Lambda a}_{\xi \{\Xi\}}$  using \eq{e5.11}.

Consider the expression in the square brackets on the r.h.s.\  of
\eq{e5.15}.  Using \eq{e5.11} one obtains a sum of terms that can be
obtained from $\prod_\xi \M^{\Lambda a}_{\xi\{\Xi\}}$  by replacing in
all possible ways the factors for some $\xi$  by  the  corresponding sums
from the r.h.s.\  of \eq{e5.11}. One can see that such a sum  can
be expressed as a sum over all sets of \UV-subgraphs
$\{\zeta\} \succ \{\xi\}$, as follows:
\be[e5.16]
\prod_\xi \M^{\Lambda b}_{\xi\{\Xi\}} -
\prod_\xi \M^{\Lambda a}_{\xi\{\Xi\}}
=
\sum_{\{\zeta\} \succ \{\xi\}}
  \prod_\xi \Z^{ba}_{\xi\{\Xi\} \{\nu\} }
  \prod_\zeta \M^{\Lambda a}_{\zeta\{\xi\}}\EOE,

\ee
where $\Z^{ba}_{\alpha_\xi\{\alpha_\zeta\}}$ depend only on
$\alpha_\zeta$ for $\zeta\subset\xi$. Substituting this result into
\eq{e5.15}, renaming $\zeta\leftrightarrow\xi $ and rearranging
the terms, we get:
\be[e5.17]
\hbox{eq.\eq{e5.13}}
=
\sum_{\{\xi\} \succ \{\Xi\}}
 \Bigl(
   \prod_\xi \Z^{ba}_{\xi\{\Xi\}}
 \Bigr)
 \cR_b \.G\{\xi\}\EOE.

\ee

Thus, we have proved that the renormalization-group transformations on
individual diagrams have exactly the same form \eq{e5.6} for our operation
$\cR$  as for the standard $R$-operation. Moreover,  we  have
established  an  explicit connection between the  variations  of  the
operation $\R$  and  the  induced variations of $\cR$. This
connection is expressed by \eq{e5.12}.

\subsection{Remarks\EOH.}
\label{ss5.6}

We would like to conclude this section with  the  following  remark.  To
prove that $\cR$  is  equivalent  to  some  other  subtraction
scheme  requires introduction of  a  regularization  that  is  natural
for  that  scheme  and commuting the limit $\Lambda\to\infty$
and the limit of taking off that regularization,  as was done in~\subsect{ss4.1} for the case of the dimensional regularization.
Such an approach is rather inelegant.

It would be more satisfactory not to reduce one  subtraction  scheme  to
another but rather to prove that the  perturbation  series  generated  by
an $R$-operation is a formal  perturbative  solution  to  some
non-perturbatively formulated equations. Such equations may need  to
require  a  more  detailed knowledge  of  the  properties  of
renormalized  diagrams; in fact, it would be
sufficient\cite{Brandt} to derive
a short distance operator product expansion. The latter is a special case
of the more general Euclidean asymptotic expansions, which
we are going to derive in momentum representation.

\section{Expansions of renormalized diagrams.}
\label{Overview}

\subsection{Formulation of the problem.}
\label{Overview.Formulation}

The feasibility and usefulness of considering asymptotic expansions
of Feynman diagrams for general Euclidean asymptotic regimes was realized
in\cite{I},\cite{Inew},\cite{II}. The formulation of the Euclidean asymptotic expansion
problem follows\cite{Inew} except that we now use GMS schemes
instead of dimensional regularization and the MS scheme (which is a special
case of GMS schemes) and consider only the simplified version of the expansion
problem without contact terms.

\paragraph*{Asymptotic regime.}
Let $G$ be an arbitrary Euclidean multiloop Feynman diagram.  Let
$G(p,\ldots)$ be its unrenormalized momentum-space integrand,
where $p$  collectively denotes its integration  (loop)  momenta  while
dots stand  for  other  dimensional parameters on which $G$ also
depends. Those parameters  include  masses  (which enter the
propagators of $G(p,\ldots)$) and external momenta, and will be
referred to as {\em external parameters\/} of the diagram $G$. We wish to
construct  asymptotic expansion for the diagram $G$  in  the
asymptotic  regime  when  some  of the external parameters of $G$ are
much larger than others.

Denote the large (or ``heavy") external parameters of $G$  collectively
as $M$, and small (or ``light") as $m$.
Formally  speaking,
\tAs-expansions that  we study require presence of a scalar parameter
with respect to which to expand.  As an abuse of notation, we will use
the same symbol $m$ to  represent  such a parameter which goes to
zero and  to  which  all  the  light  parameters are proportional.
Thus, the asymptotic regime we wish to consider is described as
$m\to 0$ and $M=O(1)$.

It makes sense to assume that neither the set $M$ nor $m$ are empty.

\paragraph*{Perfect factorization.}
An extremely important requirement on expansions at the level
of an individual diagram
is that they should ran in powers and logs of the expansion parameter.
This is a conretization of the fundamental requirement of perfect
factorization of expansions of Green functions, which means that the
coefficient functions depending on large parameters should not
contain non-analytic (logarithmic)
dependences on light parameters.
It has been realized that only expansion possessing the property of
perfect factorization have phenomenological significance; in particular,
only such expansions are useful in models with massless particles like QCD.
Moreover, such expansions possess the property of uniqueness
which turns out to be extremely useful; e.g.\ it simplifies study of gauge
properties of expansions since they then inherit the gauge properties of
non-expanded Green functions (a formalism for studying gauge properties
within the formalism of the $\tAs$-operation
was developed in\cite{pivovarov}).
Phenomenological and technical aspects of this
requirement are discussed in detail in\cite{Inew}.
Here we only note that in the papers\cite{dslav:73}
where the fact of power-and-log nature of expansions was verified
at a formal level for several asymptotic regimes,
no convenient explicit formulae for coefficient functions of expansions
were presented,
while the standard BPHZ derivation of OPE resulted in expansions in which
coefficient functions that contained all the dependence on the large
momentum also depended on the light masses in a non-trivial way.
Short distance OPE possessing the property of perfect factorization
was obtained in\cite{fvt:83}.
An important consequence was the discovery of efficient
calculational formulae for coefficient function
of OPE\cite{fvt:83},\cite{algorithm}.\footnote{
For a (attempt of) verification of some of the dimensionally regularized
results of the theory of \tAs-operation
within the framework of the (modified) BPHZ method see\cite{Smirnov:book}.
The combinatorial part of \cite{Smirnov:book} is strongly influenced by
the theory of \tAs-operation\cite{II}
but never attains transparency of the latter. On the other hand, the only
completely explicit treatment of the analytical aspects of
the theory of Euclidean asymptotic expansions remains the one given
in\cite{Tk-V:2}, \cite{KTV:fnal}, \1 and the present paper.
}

\paragraph*{Renormalization.}
We assume that the diagram $G$ (i.e.\  the integral of
$G(p, \ldots)$ over $p$ in infinite  bounds)  is  renormalized  using  the
GMS prescription.
The  GMS prescription comprises  all  the  schemes
that  possess  the property   of polynomial dependence of the
corresponding renormalization group functions on dimensional
parameters.\footnote{
The first and most important example  is the
MS scheme\cite{tHooft:MS},
for  which  the polynomiality property was established
in\cite{Collins:Ren}.
}
This is an important assumption both technically and conceptually.

Conceptually, the asymptotic expansions in masses and  momenta
obtained within such schemes  possess,  as  we  will  see,  the
above mentioned property of perfect factorization.\footnote{
The fact that the property of perfect factorization is a necessary
condition for existence of OPE in the MS scheme
was observed in\cite{analytic}.
For non-GMS schemes perfectly factorized expansions
have a more complicated form.
}

Technically, the GMS prescription amounts  to
subtraction  from the integrand of those and only those terms in the
asymptotic expansion in the \tUV\  regime that generate \tUV\  divergences,
while the necessary modification of  the logarithmic terms at zero
momentum (the operation $\r$) does  not  affect  the dependences
on the dimensional parameters. The net effect, as we will see, is to
trivialize the problem of expansion of renormalized diagrams  by
reducing it to a study of double \tAs-expansions
(see~\subsect{Overview.Limit} below).

Renormalization  introduces  an  additional  external  parameter  for
$G$ besides its masses and external momenta. Such parameter is usually
denoted as $\mu$. The dependence on $\mu$ is known explicitly: $\cR\.G$ is
 a polynomial  in  $\log\mu$.  Therefore, it is of no practical
consequence whether $\mu$ is considered as a heavy or light parameter.
For definiteness, we will treat it as a heavy parameter.

\paragraph*{External momenta as fixed parameters.}
The simplest version of the expansion problem emerges if one fixes
external momenta at some values and
then treats them on an equal footing with
masses. This is what  we will assume for now.
It should be emphasized that in principle it is not necessary
to fix the  momenta  at generic non-zero  or  otherwise
non-exceptional values as long as the initial expression one wishes
to expand is well defined. We will discuss this point in more  detail
below.  Here  we only wish  to  note  that  the precise  conditions of
when  a  diagram  is well-defined depend on the details  of  the
structure  of  specific  Feynman diagrams in a specific model,  and it
is  not  our  aim  to  discuss  such conditions. The only important
thing is that our technique is insensitive to such details.

A somewhat more complicated (and more general) version  of  the
problem would be to consider the diagram as a distribution in the
external  momenta.  Here one expects additional terms to appear in the
expansion; such  ``contact'' terms should be proportional to
$\d$-functions of linear  combinations  of  the external momenta
(cf.\   below  the discussion  of \tIR\   singularities  of  the
non-expanded diagrams). This case will be considered in a separate
publication. Note that within the framework
of dimensional regularization and the MS scheme explicit expressions
for contact terms were obtained in\cite{II}.

\paragraph*{Linear restrictions on momenta.}
Another aspect of the problem is how one divides  the  external
momenta into heavy and light. The point  is  that  certain  sums  of
heavy  external momenta should be  allowed  to  be  light,  i.e.\
$O(m)$, in  some  physically meaningful situations. This amounts
to imposing linear  restrictions  on  the heavy external momenta. Such
restrictions were analyzed in\cite{II} where an
important  class  of {\em natural restrictions\/} was  identified.
The natural restrictions can be described as follows. One divides the
external lines of $G$ through which heavy external  momenta  flow, into
 several ``bunches". Then for each bunch one imposes a single
restriction  that  the (algebraic) sum of the corresponding momenta is
$O(m)$ (i.e.\  it is equal  to  a combination of the light momenta
only).  Otherwise  the  heavy  momenta are assumed to take generic
values. Note that there always is at least one  such restriction due
to overall  momentum  conservation:  the  sum of  all  heavy momenta
must be equal to the sum of all light momenta.
As a rule our analysis at the level of individual diagrams
is quite general,
but the interpretation of the results at the level of Green functions
turns out to be more transparent for asymptotic regimes
with natural restrictions.

\paragraph*{Infrared divergences.}
It should be pointed out that Feynman integrands  in models with
massless particles possess singularities  at  finite  values  of  $p$
due  to massless propagators.
Such singularities can formally be even
non-integrable but, nevertheless,  spurious  in  the  sense  that
they cancel  out  after performing integrations and taking into
account specific algebraic properties of the integrands  like  gauge
invariance (cf.\  the  Kinoshita-Poggio-Quinn
theorem\cite{KPQ:theorem}), or they may require special  treatment
whose exact form is determined by additional  considerations.  Such
considerations are ``orthogonal" to the expansion problem  proper  (in
the  sense  explained below) and their discussion goes beyond  the
scope  of  the  present  paper.  However, the following remarks can be
made here.

One can remove such singularities from the  initial  integrand  using
a version of the special subtraction operation $\R$  introduced
in\1.  Then one can perform all the reasoning
of the present paper  taking  into account that our technique is
essentially insensitive to presence of such $\R$  in  the integrand
(see~\SUB\cref{18.4} in\1 and our~\subsect{alternative}).

Alternatively, one can regularize such singularities  by  introducing
a mass, $m_0$, for the massless particles, and after the expansion in
$m_0$ is  done, to consider  the  limit  $m_0 \to 0$.  This would  be
technically equivalent  to considering double expansion in the regime
$m_0 \ll m \ll M$. Such a problem can  be studied by
straightforward application of the results of the  present  paper.
Indeed, our main concern here is exactly the  extension  of  the
results  on simple \tAs-expansions to the case of double \tAs-expansions
(see  below~\subsect{Overview.Limit}). Needless to  say, further
extension  to  three-fold  expansions  etc.\  is completely
straightforward. The net effect is that the multiple \tAs-expansions of
 the  above  type  can  be  obtained  by  performing simple expansions
sequentially, in any order. Thus, one can first expand in the regime
when  $m_0$ and $m$ are much less than $M$, and after that, perform
termwise expansion of the result for $m_0 \ll m$. How many terms in
the expansion in $m_0$  one  should retain, and what one should do with
the singularities in $m_0$ , is to  be  decided  from the specifics of
the problem. This way to proceed is  essentially equivalent to the
first one  based  on  the use  of  the  operation $\R$,
because
the \tAs-expansion in $m_0$---as  any  \tAs-expansion---can  be
expressed  as an $\R$ with suitably chosen finite
counterterms.  However, introduction of a non-zero mass $m_0$  has an
advantage of not requiring new notations.

Either way, the factors in the final expansion that depend on the
heavy parameters of the problem (e.g.\ the coefficient functions of the
OPE  which depend on the heavy momentum $Q$) are independent of the
\tIR\  structure  of  the initial diagram.\footnote{
Another way to put it
is to say that  the  OPE  coefficient  functions  are analytic in
the light mass parameters, including the  regulator mass  $m_0$.
Therefore, whatever one does with $m_0$  afterwards (e.g.\  taking it  to
 $0$)  will not, essentially, affect the coefficient functions.
}
Only the factors in which  the  dependence  on  the light parameters is
concentrated are sensitive to \tIR\  structure. This is essentially the
property of perfect factorization.

To reiterate: what happens when $m_0$  is taken to zero depends  on
details like gauge invariance of the model and \tUV\ renormalization
procedure  adopted, and affects only those factors in the expansion
which contain the non-trivial dependence on the light parameters. On
the other hand, what we wish to focus on is the analytical
aspects of the expansion problem  proper, avoiding inessential
details. It is sufficient to say in this respect that our
techniques offers efficient ways to deal with such singularities. For
the above reasons, in what follows we will simply ignore such
singularities, in order to avoid unnecessary notational
complications. This will allow us to concentrate on the
non-trivial aspects of our techniques.

Lastly, it is convenient to assume that the set of heavy parameters
$M$ is non-empty, because otherwise the expansions in  $\k$  and
$m$  coincide and the problem degenerates into a trivial one.
(Triviality means that the renormalized diagram itself is a power
of $m$ times a polynomial in $\log m$, so that its
\tAs-expansion coincides with the diagram itself.)

\subsection{Expansion of \UV-regularized diagram.}
\label{Overview.Expansion}

A GMS renormalized diagram can be represented as
follows (cf.\ \eq{e1.4}):
\be[GMS:def]
     \cR \. G(M,m) \equiv \cR \. \int dp \, G(p,M,m)
\ee\be
     \equiv \lim_{\Lambda\to\infty} \int dp \, \Cut^\Lambda(p)
     \bigl[
        G(p,M,m) - \rAs \. G(p, \k M, \k m)
     \bigr]_{\k=1} .
\ee
We have introduced an auxiliary parameter $\k$ on the r.h.s.\  of
\eq{GMS:def} in order to formally describe the asymptotic expansion of
the integrand  in the  \tUV\  regime in which all dimensional
parameters of $G$ are small as compared to  the cutoff $\Lambda$.

What we ultimately wish to do is to determine  the  explicit
expression for the \tAs-expansion of \eq{GMS:def} in $m\ll M$
which we denote by  the same  symbol $\As$  as we use for the
\tAs-operation on non-integrated products:
\be[Ass.R:descr]
     \cR \. G(M,m) \asy{m}{0} \As_m\.\cR\.G(M, m) = \: ?
\ee
Such a notation is very natural both in view of the  general
definitions of section~\cref{15} of\1 and because the new version
of the operation $\As$ \eq{Ass.R:descr}, as we will see, is closely
related to the operation $\As$ already defined on non-integrated
products. Note that the term ``\tAs-operation'' was first introduced
for {\em integrated\/} diagrams\cite{II}.

It is natural to try to find the expression for \eq{Ass.R:descr} by
applying the operation $\Asm$, which has  already  been  defined  on
products of singular functions, to the expression in square brackets
on the r.h.s.\  of \eq{GMS:def}.  As we will see, this is indeed
possible. Moreover, the corresponding calculation exhibits a recursive
pattern: in order to derive \eq{Ass.R:descr} for $G$ itself one has to
assume validity of  an expansion  of  the type  \eq{Ass.R:descr}  for
integrated renormalized graphs with a lesser number of loops than $G$.

\paragraph*{Studying eq.\eq{GMS:def} at fixed $\Lambda$.}
First of all, let us fix $\Lambda < \infty$ and study
the expansion at $m\to 0$  of
the resulting  ``regularized"  integral.  One  can  immediately
write down   the expansion for  the  first  term  on  the  r.h.s.\   of
\eq{GMS:def},  $\int dp\, \Cut^\Lambda(p) \, G(p,M,m)$,
by directly using
the techniques of\1, i.e.\   by  applying  the
\tAs-operation to $G(p,M,m)$:
\be
     \int dp\, \Cut^\Lambda(p) \, G(p,M,m)
         \asy{m}{0}
     \int dp \, \Cut^\Lambda(p) \, \As_m \. G(p,M,m).
\ee
In order to expand the remaining contributions  to  \eq{GMS:def},  one
should consider the structure  of  the  expression  $\rAs\.G$  in  more
detail.  From eqs.\eq{e2.22}--\eq{e2.24} it follows
that:
\be[rAs:def]
     \rAs \. G(p, \k M, \k m)\bigr|^\0_{\k=1} \kern8cm
\ee\be
     = \sum_{\gamma < G} \sum_\alpha
     \R \.
     \bigl\{
           \d_\alpha(p_\gamma) \,
           \Tk \. G\backslash\gamma (p, \k M, \k m)
     \bigr\}_{\k=1}
     \times
     \cR \. \int dp'_\gamma \, \P_\alpha(p'_\gamma) \,
     \gamma(p'_\gamma, M, m).
\ee
(Note that the summation  here  runs  over subgraphs  corresponding  to
the singularities of the formal expansion in $\k$.\footnote{
We will also
have to deal with  singularities  of,  and the  corresponding
operation $\R^m$ for, the expansion in $m$. The resulting
notational  complications will be dealt with
in~\subsect{Structure.Defs} below.})

As a convenient abuse of notations we will often
omit $\k$ in expressions
like the r.h.s.\ of \eq{rAs:def} everywhere except in $\Tk$:
\be[not-k]
     \Tk \. G\backslash\gamma (p, \k M, \k m)_{\k=1}
     \to \Tk \. G\backslash\gamma (p, M, m).
\ee
This should cause no ambiguity.

The r.h.s.\  of \eq{rAs:def} is a sum of terms in which the
dependences  on the integration  momenta  $p$   and  on  the  external
parameters  are completely factorized. Indeed, the distribution
$
\R\.\bigl\{
       \d_\alpha(p_\gamma) \,
       \Tk \. G \backslash \gamma (p, \ldots)
    \bigr\}
$
is a polynomial of the external parameters $M$ and $m$ (this is
due to the action of $\Tk$ and the fact that
$\R$  does  not affect   $M$- and  $m$-dependences).
On the other hand, the factor
$
     \cR \. \int dp'_\gamma \, \P_\alpha(p'_\gamma) \,
     \gamma(p'_\gamma, M, m)
$
contains a non-trivial dependence on $M$ and  $m$  but is
independent of $p$. Moreover, the latter factor has the same  form  as
the initial expression
\eq{GMS:def} up to a replacement $p \to p_\gamma$  and  $G(p) \to
\P_\alpha(p_\gamma) \times \gamma(p_\gamma)$.  Therefore, it is natural to make an
inductive assumption that the operation $\As$  has been defined on the
products
$
     \cR \. \int dp'_\gamma \, \P_\alpha(p'_\gamma) \,
     \gamma(p'_\gamma, M, m)
$
for  all  $\gamma<G$. Then  the \tAs-expansion of \eq{rAs:def} is
given by applying such $\As$  to the
last  factor  in \eq{rAs:def}:
\be[Ass.rAs:def]
     \rAs \. G(p, M, m)
         \asy{m}{0}
     \As_m \. \rAs \. G(p, M, m)
\ee\be
     =
     \sum_{\gamma < G} \sum_\alpha
     \R \.
     \bigl\{
         \d_\alpha(p_\gamma) \,
         \Tk \. G\backslash\gamma (p, M, m)
     \bigr\}
     \times
     \As_m \. \cR \. \int dp'_\gamma \, \P_\alpha(p'_\gamma) \,
     \gamma(p'_\gamma, M, m).
\ee
It is important to note that there has emerged a  compact  recursive
pattern which is characteristic of our techniques: expansion of the
GMS renormalized diagram $G$ \eq{GMS:def} is reduced to an essentially
similar problem  but  with  a lesser number of integration momenta.

In the case of a single loop momentum in the initial diagram  $G$
(or one-dimensional $p$  if one does not limit the discussion
to Feynman diagrams proper), our expansion problem degenerates into a
trivial  one.  Because  of this and owing to the explicit recursive
pattern in \eq{Ass.rAs:def} we can assume that the problem has been
solved for all GMS-renormalized diagrams with lesser number of loop
momenta, i.e.\  that the operation $\As$ on the r.h.s.\   of
\eq{Ass.rAs:def} is well defined.  Using this
inductive assumption completes the  expansion
of the regularized integral.

Explicit expressions for $\As_m\.\cR\.G$ which provide solution to
the above recursive procedure, will be presented below
in section~\sect{Structure}.

\subsection{The limit $\protect\Lambda \protect\to \protect\infty$.}
\label{Overview.Limit}

The only important question of analytic nature
that needs to be  answered  is  whether
the asymptotic expansion constructed above for fixed $\Lambda$ remains such
after taking the limit $\Lambda\to\infty$, i.e.\  whether the
\tAs-operation defined by the expression\footnote{
As should be clear from the above construction,
the composition of the  two
\tAs-operations here is purely  algebraic:  the second  operation  is
applied termwise  to  the  series  generated  by the  first  one
irrespective   of approximation properties of the resulting
expression.
}:
\be[Ass.R:def]
     \As_m \. \cR \. \int dp \, G(p,M,m)
     \bydef
     \lim_{\Lambda\to\infty}
     \int dp \. \Cut^\Lambda(p)
     \As_m \. \bigl[1-\rAs\bigr] \. G(p, M, m)
\ee
delivers a true \tAs-expansion for the integral \eq{GMS:def}. The
answer is yes, and it can be justified in two steps.

\textindent{({\it i\/})}
As a first step, it is natural to consider existence of the limit on
the r.h.s.\  of \eq{Ass.R:def}. To this end we split the integration
region  in \eq{Ass.R:def} into two parts by introducing an
intermediate cut-off  at  the radius  $\mu$  in order to explicitly
separate the non-trivial asymptotic region $p\to\infty$ from the
point $p = 0$ where the expression is complicated by the operator
$\r$ but the expansion is essentially straightforward:
\be[Ass.R:split]
     \As_m \. \cR \. \int dp \, G(p,M,m)
     \equiv
     \int dp \, \Cut^\mu(p) \, \As_m \. \bigl[1-\rAs\bigr] \.
     G(p, M, m)
\ee\be
\kern4cm
     +
     \lim_{\Lambda\to\infty}
     \int dp \, \Cut^{\Lambda}_{\mu}(p)
     \As_m \. \bigl[1-\Aspk\bigr] \. G(p, M, m).
\ee
(For definition of the functions $\Cut$ see~\SUB\cref{8.5} of\1.)
The second term in this expression will be finite  if  the  two
\tAs-operations commute:
\be[Ass.Aspr:comm]
     \As_m \. \Aspk = \Aspk \. \As_m.
\ee
Then the operator $1-\Aspk$ can be taken out  to  the  left  of
$\As_m$, so  that existence of the limit $\Lambda\to\infty$ will follow
automatically (recall in this respect the motivations
and construction of the operation $\cR$ in section~\sect{R.UV.Motivations}).

The commutativity \eq{Ass.Aspr:comm} (see also  \eq{rcom})
is one of the central results of the present paper.
Its nature is essentially algebraic. Here we only note that
explicit formulae for $\As\.\cR\.G$ follow from \eq{rcom} (see
section~\sect{Structure}).

\textindent{({\it ii\/})}
The second step is analytic in  nature:  one  verifies  that  the
remainder of the expansion \eq{Ass.R:split} vanishes at the required
rate as $m\to 0$. This is formally expressed as
\be[(1-Ass).R:def]\relax
     \bigl[1-\As_m^n\bigr] \. \cR \. \int dp \, G(p,M,m)\kern7.5cm
\ee\be
     \kern.8cm\bydef \lim_{\Lambda\to\infty}
     \int dp \, \Cut^\Lambda(p)
     \bigl[1-\As_m^n\bigr] \.\bigl[1-\rAs\bigr] \.
     G(p, M, m) = o(m^n).
\ee
To check \eq{(1-Ass).R:def}, one splits the integration region as in
\eq{Ass.R:split}:
\be\relax
     \bigl[1-\As_m^n\bigr] \. \cR \. \int dp \, G(p,M,m) \kern5cm
\ee\be
     \quad\quad \equiv
     \int dp \, \Cut^\mu(p) \bigl[1-\As_m^n\bigr] \. \bigl[1-\rAs\bigr] \.
     G(p, M, m)
\ee\be
     + \lim_{\Lambda\to\infty} \int dp \, \Cut^{\Lambda}_{\mu}(p)
     \bigl[1-\As_m^n\bigr] \. \bigl[1-\Aspk\bigr] \. G(p, M, m).
\ee
For the first term on the r.h.s.\  the estimate \eq{(1-Ass).R:def} is
true by definition of $\As$ and because $\Cut^\mu(p)$ is an ordinary test
function.  For the second term,  one represents $\Cut^\Lambda_\mu(p)$ as
an integral over spherical layers of radius $\lambda$
(see~\SUB\cref{8.5} of\1):
\be[radial]
     \Cut^\Lambda_\mu(p)
     = \int_{\mu}^{\Lambda} \frac{d\lambda}{\lambda} \,
       \cut_\lambda(p).
\ee
Then one rescales the integration variable $p \to \lambda p$  and  uses
the uniformity properties of $G$ to arrive at the following expression:
\be[n1.11]
     \int_{\mu}^{\Lambda} \frac{d\lambda}{\lambda}
     \int dp \, \cut_1(p) \, \bigl[1-\As_m^n\bigr] \.
     \bigl[1-\As_{\k/\lambda}\bigr] \. G(p, M\k/\lambda, m\k/\lambda).
\ee
Recall that one can retain only those terms in $\Aspk$ that are
responsible for \tUV\  divergences (see the text immediately after
eq.\eq{e1.4}). Since $\lambda$ always divides $\k$
in the above expression, one can see that the
$o(m^n)$ estimate for \eq{n1.11} follows from  an  estimate of the
type
\be[Double.Est]
     \Bigl| \int dp \, \cut_\lambda(p) \, \bigl[1-\As_m^n\bigr] \.
     \bigl[1-\As_\k^l\bigr] \. G(p, \k M, \k m) \Bigr|
     < o(m^n) \times o(\k^l).
\ee
This can be  adopted  as  (part  of)  an  exact
analytic  interpretation
of the informal statement that the algebraic
composition of the  two  (commuting)
\tAs-operations yields a true double asymptotic expansion in the sense
of distributions  for the integrand $G(p, \k M, \k m)$.

The inequality \eq{Double.Est} is easy to understand at a heuristic level.
Indeed, the remainder of an asymptotic expansion is often estimated
(at least in the cases when the expansion
has a relatively simple analytic nature---as in our case
where one deals with expansions of integrals of rational functions,
however cumbersome)
by the last discarded term, which in our case is
$O(m^{n+1})\times O(\k^{l+1})$ (up to inessential logarithms).
This immediately explains \eq{Double.Est}.

Actually, the inequalities  that  we  prove  in  Theorem  1  (see
below section~\sect{Double}) are more stringent than \eq{Double.Est};
in particular, they also describe dependence of the bounds on the support
of the test function. This is needed in order to carry on the induction.

\subsection{Summary}

We have exhibited  the  recursive  structure  of  the \tAs-operation
for GMS renormalized diagrams and identified the inductive assumptions
as well as the propositions that have to be proved.
To proceed, we
first have to  study the structure of our expressions  in more  detail
and  derive an  explicit expression for $\As_m\.\cR\.\int dp\,G$.
This will allow us to  check  the formal commutativity of the two
\tAs-operations $\As_m$  and $\As_\k$.
Second,  we have to present and
prove the inequalities that guarantee validity of \eq{Double.Est}.

\section{Explicit expressions for $\protect\Asm\.\protect\cR\.G$.}
\label{Structure}

We are now going to derive explicit formulae for \eq{Ass.R:descr}.
We will do this assuming (by induction) that
the two $As$-operations commute  on subgraphs of $G$.
After  that  it  will not  be  difficult to check the commutativity
on $\G$ itself.
Note that the recursion  is  correct because  whatever property one wishes to
prove for a graph, one only has to make assumptions about its subgraphs.

\subsection{Some notations.}
\label{Structure.Defs}

\paragraph*{Operation $\protect\cR$ on integrands.}
So far we have been using the operation $\cR$ defined
on integrated diagrams. But since now we will have to work with
integrands, it is convenient to use the same symbol $\cR$ to denote
the operation of \tUV\  subtractions {\em prior\/} to integrations over $p$:
\be[cRdef]
     \cR \bydef 1-\rAs.
\ee
This operation is defined on graphs $G$ as well as on the products of the
form
$\P_ a(p_\gamma)\times\gamma(p_\gamma)$
where $\P_ a(p_\gamma)$  is a polynomial while $\gamma$ is any ($\k$- or
$m$-) subgraph of $G$.

\paragraph*{$\protect\cR$ associated with $\protect\R$.}
Although the  entire  arbitrariness  in  the  definition  of $\cR$ for
an individual diagram $G$ is in the operator $\r$
(more precisely, $\r_{(G)}$), it is  more
convenient  to think about $\r$ in terms of the  operation
$\R$ (which also involves operators $\r_{(\gamma)}$ for $\gamma<G$).
This is because  all  the explicit expressions for $\cR$
(cf.\  eq.\eq{e2.22}) involve $\R$.
On the other hand, the two points of view are equivalent if one recalls
that in  the problems  of perturbative quantum field theory one deals with
the entire universum of graphs: specifying the family of operations
$\R_{(G)}$ on the entire universum of graphs $G$ is equivalent to specifying
the family of operators $\r_{(G)}$. Then, to fix an operation $\cR$
(or, equivalently, $\rAs$) on a
hierarchy of graphs, it is sufficient to fix an operation $\R$ on it.

For the above reasons we will say that an operation $\cR$ (and
$\rAs$) is
{\em associated with\/} some operation  $\R$,  whenever  it  is
necessary  to  indicate this kind  of relationship between the two
operations.

\paragraph*{$\k$- and $m$-subgraphs.}
Since there are two expansions---in $\k$ and $m$---in our
problem that one has to deal with simultaneously,
there are two systems of singular planes, complete subgraphs
etc.\  in the same graph  $G$.\footnote{
In general, one should
also  consider  the third  system of  subgraphs---the one
corresponding to singularities of unrenormalized non-expanded
integrand $G(p,\ldots)$
that were discussed in~\subsect{Overview.Formulation}.
In order to keep notations simple, we agreed not  to  indicate
explicitly   the possible presence   of such singularities, the
corresponding operation $\R$ etc.
}
In order to distinguish the objects from the two systems,
we  will call them $\k$- and $m$-subgraphs etc.

We will retain the symbols $\R$ and $\cR$ for the operations associated
with $\k$-singularities, and use the symbols $\R^m$ and $\cR^m$
for the operations associated  with $m$-singularities.

By default, a subgraph in a formula is $\k$-subgraph.
Presence of $m$-subgraphs will be explicitly indicated.

The relation between the two systems of subgraphs is based on a
simple principle: any factor $g\in G$ which is ``$m$-singular"
(i.e.\ develops singularities after expansion in $m$)
is automatically ``$\k$-singular"---because the set of momenta
and masses with respect to which the expansion is done in the latter case
comprises all such parameters in the former case.
Thus, to every $m$-subgraph $\Gamma$
there corresponds a unique $\k$-subgraph $H$ obtained from $\Gamma$
by ``$\k$-completion".

\subsection{General formula for the expansion $\As_m\.\cR\.G$.}
\label{Structure.Explicit}

Let us first explain the structure of \eq{Ass.Aspr:comm}.
The dependence on $p$, $m$ and $M$ in \eq{Ass.Aspr:comm}
can be described as follows:
\be[ccc1]
     \Aspk \. G(p, M, m)
     = \sum D(p) \times C( M, m),
\ee
which correponds to the expansion \eq{rAs:def}. It is clear that
$D(p)$ are distributions well-defined everywhere except for the point
$p=0$.
The action of $\Asm$ in \eq{Ass.rAs:def} can be described as
\be[ccc1.1]
     \Asm \. \Aspk \. G(p, M, m)
     \equiv \sum D(p) \; \Asm \. C( M, m)
     = \sum D(p) \; A( M) \; B( m)
\ee
(the explicit fromulae for $\Asm\.C$ are yet to be determined).
On the other hand,
\be[ccc2]
     \Asm \.  G(p, M, m)
     = \sum C'(p, M) \; B'( m)
\ee
(cf.\ the explicit expression below in \eq{Ass:def}), and
\be[ccc3]
     \Aspk \. \Asm \.  G(p, M, m)
     \equiv  \sum \Aspk \. C'(p, M) \; B'( m)
     = \sum D'(p) \; A'( M) \; B'( m).
\ee
So far we don't know the form of $C'$ and how $\Aspk$ does its job on $C'$
(which will be explained below). Nevertheless,
the commutativity \eq{Ass.Aspr:comm} implies that $D=D'$ etc.\footnote{
Strictly speaking $D$ etc.\ depend on a summation index so that
one may need to take linear combinations
to establish the equality.
}
It follows immediately that the
commutativity is preserved if one replaces $\Aspk$ by $\rAs$:
\be[rcom]
     \Asm \. \rAs = \rAs \. \Asm.
\ee
Indeed, on both sides of \eq{rcom} the operator $\r$
acts---as demonstrated above---on exactly the same $p$-dependent
distributions $D(p)$.

The explicit version of \eq{ccc2} is analogous to \eq{rAs:def}
(cf.\ eqs.\eq{e2.22}--\eq{e2.24}):
\be[Ass:def]
     \Asm\.G(p,M,m)
     =
     \sumx {\emptyset\leq\Gamma<G} {$\Gamma$ is $m$-subgraph}
     \sum_ a
         \R^m \. \bigl[ \d_ a(p_\Gamma) \, \Tm
                        \. \GG (p,M,m)
		 \bigr]
     \times
         \cR^m \. \langle \P_{ a,\Gamma} * \Gamma \rangle ,
\ee
where
\be[angled]
     \cR^m \. \langle \P_{ a,\Gamma} * \Gamma \rangle
   \equiv
     \langle \P_{ a,\Gamma} * \cR^m \. \Gamma \rangle
   \equiv
     {\displaystyle\lim_{\Lambda\to0}}
     \int dp'_\Gamma \, \Cut^\Lambda(p) \,
     \P_{ a,\Gamma} (p'_\Gamma) \, \cR^m \. \Gamma (p'_\Gamma,m).
\ee
Note that since $\Gamma$ is  an  $m$-subgraph  of  $G$, the  object
\eq{angled} is an almost-uniform (in the sense
of\1, subsec.1.2) function of $m$ and  is independent of
$M$ and $p$. On the other hand, all the dependence on $p$  is
concentrated  in the square-bracketed factor on the r.h.s.\   whose
$m$-dependence  is trivial  while  the $M$-dependence is not.

We now wish to apply the operation $\cR=1-\rAs$ termwise
to the above expansion (similarly to \eq{ccc3}).
Owing to \eq{rcom} this is equivalent to $\Asm\.\cR\.G$.
Since \tAs-expansion of a product is a product of
\tAs-expansions\1 and since
the only non-trivial dependence on $\k$ is via $M$ in
$\GG (p,M,m)$, one has:
\be[R.Ass:def]\label{General.Formula}
     \Asm\.\cR\.G(p,M,m) \kern 7cm
\ee\be
     =
     \sumx {\emptyset\leq \Gamma < G} {$\Gamma$ is $m$-subgraph}
     \sum_a
        \cR \. \R^m \.
	\bigl[
           \d_{ a,\Gamma} (p_\Gamma) \, \Tm \. \GG (p,M,m)
        \bigr] \,
     \times
        \cR^m \. \langle \P_{ a,\Gamma} * \Gamma \rangle.
\ee
The action of $\cR$ on the somewhat unusual
expression in square brackets will be explained
in the next subsection.

The above eq.\eq{R.Ass:def} constitutes the final result of the analysis
of the problem of Euclidean asymptotic expansions of Feynman diagrams
as seen from the point of view of the abstract theory of \tAs-expansions
of products of singular functions. Below in section~\sect{Diagrammatic}
we will transform it---using
specific properties of Feynman diagrams proper---to a more convenient
form similar to the \tAs-operation as presented in\cite{II}.
The immediate remarks to be made here are as follows.

\textindent{({\it i\/})} The non-analytic dependences on the heavy parameters $M$ and the light
parameters $m$ are clearly factorized in \eq{R.Ass:def}.
Indeed, the expression in angle brackets is independent of $M$ by construction.
On the other hand, the $M$-dependent distributions over $p$
in square brackets are
pure power series in $m$ (due to the action of $\Tm$ and the fact that
neither $\R^m$ nor $\cR$ affect the resulting powers of $m$).
Therefore, it is clear---in the context of ordinary
short-distance OPE---that the angle-bracketed expressions correspond
to matrix elements of OPE (the polynomials $\P$ then correspond to vertices
with composite operators)
while the square-bracketed expressions, to coefficient functions.

\textindent{({\it ii\/})} All the quantities in \eq{R.Ass:def} are finite by construction:
the angular-bracketed ``matrix element" has its \tUV\  divergences removed
by $\cR^m$, while the \tIR\  and \tUV\  singularities of the square-bracketed
``coefficient functions" are eliminated by $\R^m$ and $\cR$, respectively.

\textindent{({\it iii\/})} The expression \eq{R.Ass:def} as a whole is independent
of the specific choice of the operation $\R^m$ and the associated
operation  $\cR^m$. This is a usual feature of representation
of an \tAs-expansion in terms of an intermediate $\tilde{R}$-operation
(recall that an \tAs-expansion is unique\cite{Inew},\cite{Volume.I}).

\subsection{Defining $\rAs\.\R^m\.[\ldots]$ in \protect\eq{R.Ass:def}.}

Consider the following object from \eq{R.Ass:def} ($p'_\Gamma$ is the integration
momentum implied in $\langle\ldots\rangle$; it is isomorphic to $p_\Gamma$):
\be[(3.1)]
     \sum_a
        \d_a(p_\Gamma)\,
        \Tm\.\GG(p,M,m)
     \times
        \P_{a,\Gamma}(p'_\Gamma)
     \quad\quad\hbox{($\Gamma$ is $m$-subgraph)}.
\ee
When $\d_a$ is integrated out, its derivatives also affect $\GG$.
Let us exhibit this explicitly using the fact that the above pattern of
the polynomials $\P_a$ and  the $\delta$-functions $\d_a$ is characteristic
of the operation of Taylor expansion (cf.\ eq.(\cref{9.6})  in\1).
We use the following elementary identity:
\be[(3.2)]
     \sum_a \d_a(p_\Gamma) F(p_\Gamma) \times \P_a(p'_\Gamma)
     =
     \sum_b \d_b(p_\Gamma) \times \T_{p'_\Gamma} \. F(p'_\Gamma) \, \P_b(p'_\Gamma) .
\ee
Now split the variable $p$ as
\be
     p = ( p_\Gamma, p_{\GG}),
\ee
and define
\be[(3.3)]
     m_\Gamma = (m,p_\Gamma),\kern1.3cm m'_\Gamma = (m,p'_\Gamma).
\ee
Then we can rewrite \eq{(3.1)} as
\be[(3.4)]
     \hbox{eq.\eq{(3.1)}}
     =
     \sum_b \d_b(p_\Gamma) \times
     \bigl[
         \Tmp \. \GG(p_{\GG}, M, m')
     \bigr]
     \, \P_b(p'_\Gamma) .
\ee
The variables $p_{\GG}$ on the r.h.s.\ parameterize the singular plane
$\pi_\Gamma$ on which the entire expression is localized. Note that
$\GG$ is being expanded in $p'_\Gamma$, so that all the singularities
are with respect to $p_{\GG}$.

The above formula allows one to define $\R^m$ on such expressions
(cf.\ the reasoning in~\Sect{rf.Definition} above
and~\SUB\cref{14.2} of\1)) as follows:
\be[(3.5)]
     \sum_a
          \R^m \. \bigl[
               \d_a(p_\Gamma) \, \Tm\.\GG(p,M,m)
	  \bigr]
          \times \P_{a,\Gamma}(p'_\Gamma)
          \quad\quad\hbox{($\Gamma$ is $m$-subgraph)}
\ee\be
     \bydef
     \sum_b \d_b(p_\Gamma) \times
     \R^m \. \Tmp \. \GG(p_{\GG}, M, m') \, \P_b(p'_\Gamma)
\ee\be
     =
     \d(p_\Gamma) \times \R^m \. \Tmp \. \GG(p_{\GG}, M, m')+\ldots,
\ee
where on the r.h.s.\ we have not shown the terms proportional to derivatives of
the $\d$-function---such derivatives vanish after integration over $p$ which
eventually has to be done.
Note that
there need not be any correlation between the definitions of the operation
$\R^m$ on $\Tm G$ and on $\Tmp\.\GG$
(cf.\ the reasoning in section~\sect{rf.Definition}).

In \eq{R.Ass:def}, the expression \eq{(3.5)} is being acted on by $\cR=1-\rAs$.
Here one only has to deal with the non-trivial dependence on $M$ on the r.h.s.,
and it is easy to understand that $\Aspk$ should be applied termwise to the
r.h.s., effectively getting combined with $\R^m$.\footnote{
In\1 the formulae for the \tAs-operation were presented only for a class
of singular functions without non-integrable singularities prior to expansion.
In the present case, we for a first time encounter a situation where an
\tAs-operation---$\Aspk$---is being applied to an expression
\eq{(3.5)} which involves an $\tilde{R}$-operation.
As was noted in\1, extension of the formula for
\tAs-operation to the case of a singular initial expression is
straightforward (see also below~\subsect{alternative}).
}
It remains to note that, as usual,
the operator $\r$ on $\Aspk\.\R^m\.\Tmp\.\GG$ (or, equivalently, $\R$ on the
$\Tk\.\Tmp\.\GG\equiv \T_{\k'}\. \GG$, where $\k'=(\k,p'_\Gamma)$) can be chosen
independently from $\r$ on $\Aspk\.\R^m\.\Tm\.G$
(cf.\ section~\sect{rf.Definition}).

Finally, eq.\eq{R.Ass:def} takes the form:
\be[(1.9n)]
     \Asm\.\cR\.G(p,M,m) \kern9cm
\ee\be
     =
     \sumx {\emptyset\leq \Gamma < G} {$\Gamma$ is $m$-subgraph}
     \d(p_\Gamma) \times \cR\.\int dp'_\Gamma \R^m\.\Tmp\.\GG(p_{\GG}, M, m') \,
     \cR^m\.\Gamma(p'_\Gamma,M,m)
     +\ldots.
\ee
Note that $\GG$ depends on $p'_\Gamma$ through $m'$. It should also be remembered
that the integration over $p'_\Gamma$ should be understood in the sense of the principal
value (operation $*$).

\Eq{(1.9n)} eliminates the last unknown in \eq{R.Ass:def} and represents a
convenient starting point for studying exponentiation of \tAs-operation on collections
of Feynman diagrams corresponding to Green functions
(section~\sect{Diagrammatic} below).

\subsection{\tAs-operation on products involving $\R$ and $\d$-functions.
Proof of commutativity of the two \tAs-operations.}
\label{alternative}

Both in \eq{R.Ass:def} and in \eq{(1.9n)} one has to consider
\tAs-expansions of products involving an $\tilde{R}$-operation
and/or $\d$-functions.
Therefore, it is worthwhile to consider this point from a general point of
view. Moreover, it turns out that the object
\be[object]
     \R^m \.
	\bigl[
           \d_{ a,\Gamma} (p_\Gamma) \, \Tm \. \GG (p,M,m)
	\bigr]
\ee
from \eq{R.Ass:def} can be analyzed
without performing projections onto the plane singled out by the
$\d_{ a,\Gamma} (p_\Gamma)$---in complete analogy with
expressions of the form $\R^m\.G$. The simplicity and straightforward character
of the resulting formal proof of commutativity \eq{rcom}
is another example of how
the meticulous attention to the formalism and notations
in the preceding paper\1 pays off.

Indeed, with some experience with the formalism
and understanding of the mechanism of the \tAs-operation, an
explicit expression for $\cR$ on the distribution in square brackets
in \eq{R.Ass:def} can be written down offhand. In view of
\eq{cRdef}, it is sufficient to present expressions for the operation
$\rAs$.

To start, recall the expression for $\rAs\.G$ \eq{rAs:def}, which we here rewrite
in slightly different notations similar to those used above:
\be[rAs.G]
     \rAs\.G(p,M,m)
     = \sumx {\emptyset\leq\gamma<G} {$\gamma$ is $\k$-subgraph} \sum_b
     \R \. \bigl[
          \d_b (p_\gamma) \, \Tk \. G(p,M,m)
	   \bigr]
     \times \cR \. \langle \P_b * \gamma \rangle .
\ee
First we
wish to write down a similar expression for the product $G$ replaced as
\be[G']
     G(p,M,m)\to G'(p,M,m)
     =
     \R^m \. \Tm \. G (p,M,m).
\ee
To put it simply, some of the factors have been replaced
(owing to the action of
$\Tm$) by arbitrarily singular factors; on top of everything, the
singularities of the resulting expression have been subtracted using $\R^m$.
Expressions of this form appear in \eq{(1.9n)}.

Second, we also wish to consider expressions obtained from \eq{G'}
by replacing  the group of factors corresponding to the
$m$-subgraph $\Gamma$ by one factor---the $\d$-function $\d_a$:
\be[G"]
     G'(p,M,m)\to G''(p,M,m)
     =
     \R^m \. \bigl[
          \d_a (p_\Gamma) \, \Tm \. \GG (p,M,m)
     \bigr].
\ee
The crucial thing to realize is that neither the philosophy nor the reasoning
of the theory of \tAs-operation developed in\1 need to be changed in order
to deal with such products. Indeed,
as to the singularities and the
additional $R$-operation (the operation $\R^m$ in our case),
one only has to bear in mind the following.
As a ``formal expansion" one should take the original product {\em without\/} the
additional $R$-operation but with the factors that can be expanded, expanded.
One then proceeds to constructing the counterterms (introducing an
intermediate $\tilde{R}$-operation etc.) treating on an equal footing both
the ``old" singularities
(i.e.\ corresponding to the factors that are singular prior to expansion)
and those generated by formal expansion. However, in constructing the
counterterms via consistency conditions, one uses the initial expression which
contains {\em both\/} non-expanded factors {\em and\/} the additional $R$-operation.
The entire procedure becomes perfectly obvious if one recalls the philosophy
of constructing the expansion by considering it first in
an open region in the space
of $p$ where all the factors are regular and then expanding the domain of
definition of the expansion by adding counterterms proportional to
$\d$-functions. By analogy with \eq{rAs.G},
one immediately obtains a similar expression for $G'$:
\be[rAs.G']
\herring{
\=    \rAs \. \R^m \.
          \Tm \. G(p,M,m)
\\
\> \quad = \sumx {\emptyset\leq\gamma<G} {$\gamma$ are $\k$-subgraphs} \sum_b
     \R \. \bigl[
          \d_b(p_\gamma)\Tk \. G \backslash \gamma (p,M,m)
	   \bigr]
     \times \cR \. \langle
          \P_b * \R^m \. \Tm \. \gamma (p,M,m)
     \rangle.
}
\EOE{}

\ee
Note that we have replaced $\Tk\.\Tm$ acting on $G\backslash\gamma$ by $\Tk$.

Turning to the expression \eq{G"}, it is not difficult to realize
that the $\d$-function is as good as any other
factor from the point of view of the formalism---provided one considers it as
a singular factor. Its only effect is that now
every $\k$-subgraph $h$ must contain the $\d$-function
plus, perhaps, some other $\k$-singular factors.
A simple combinatorial observation is
that the subgraphs $h$ are in one-to-one correspondence
with $\k$-subgraphs $\gamma$ of $G$ such that $\gamma\ge\Gamma$.
Without more ado we get:
\be[rAs.G"]
\herring{
\=   \rAs \. \R^m \. \bigl[
          \d_a (p_\Gamma) \, \Tm \. G \backslash \Gamma (p,M,m)
     \bigr]
\\
\>\qquad  = \sum_{\Gamma\leq\gamma<G} \sum_b
     \R \. \bigl\{
          \d_b (p_\gamma) \, \Tk \. G \backslash \gamma (p,M,m)
     \bigr\}
\\
\>\qquad\qquad  \times \cR \. \langle
          \P_b * \R^m \. \bigl[
               \d_a (p_\Gamma) \, \Tm \. \gamma\backslash \Gamma (p,M,m)
	  \bigr]
     \rangle.
}
\EOE{}

\ee
One should remember that in this expression $\gamma$ are $\k$-subgraphs in $G$,
while $\Gamma$ are $m$-subgraphs. Summation runs over $\gamma$ in accordance with the
above observation about one-to-one correspondence between $\k$-subgraphs
in $G$ and in the product $G''$ given by \eq{G"}.

It is a good excercise to verify equivalence of \eq{rAs.G"} and the expressions
that can be obtained along the lines of the preceding subsection.

It remains to note that now one can easily check the formal commutativity of
the two $As$-operations \eq{rcom} by substituting \eq{rAs.G'} into the
r.h.s.\ of \eq{rcom} and using the expression for its l.h.s.\ that has already
been discussed---provided $\Asm$ on the l.h.s.\ can be put under integration.
The final justification for the latter comes from inequalities of the type
\eq{Double.Est} that will be obtained in section~\sect{Double}.

\section{Diagrammatic form of $\protect\Asm\.\protect\cR\.G$.}
\label{Diagrammatic}

The formulae derived in the preceding section are not immediately useful for
deriving \tAs-expansions for Green functions in OPE-like form.
To this end one should recast eq.\eq{(1.9n)} into a form that would take
into account specific properties of Feynman diagrams that were irrelevant
at the analytic stage.

The diagrammatic analysis of Euclidean \tAs-expansions was performed
in much detail in\cite{II}.
The combinatorial aspects which we are going to discuss in this section
are not very
sensitive to whether one deals with expansions in dimensionally regularized
form, as in\cite{II}, or in formalism without regularizations, as in
the present paper.
Therefore, we will give only an outline of the reasoning
and consider just two key examples: the ordinary short-distance OPE and the
expansion in heavy masses (which extends the decoupling theorem of
Appelquist and Carrazzone---for a review see\cite{Collins:Ren}).
An interested reader can find further examples and a detailed description of
the combinatorial techniques in\cite{II}.

\subsection{Using factorization properties of $\GG$.}

At this point we may suppose that $G$ is an ordinary Feynman diagram
and use the factorization properties of the expression
$\GG(p, M, m)$.\footnote{\ignorespaces
The reasoning below follows section~\sect{rf.Definition}
which in turn is reminiscent of\cite{II}.
}

An $m$-subgraph $\Gamma$ is any set of lines and vertices of $G$  such that
every line from $\Gamma$ is singular after expansion in $m$ (irrespective of
whether or not the line is singular before expansion); $\Gamma$ must also
satisfy the  completeness condition (see~\SUB\cref{7.2} in\1).
In the present case completeness of $\Gamma$ means the following:
\textindent{({\it i\/})} When one nullifies all light external
momenta from the set $m$ as well as all the momenta flowing through
the lines of $\Gamma$, no other $m$-singular line  of $G$  will have its
momentum nullified owing to momentum conservation at vertices;
\textindent{({\it ii\/})} $\Gamma$ contains all those  and  only  those
vertices of $G$  whose all incident lines belong to $\Gamma$; no heavy
external momentum from  the  set $M$  is allowed to enter into such
a vertex.

Consider the complement of $\Gamma$ in $G$, denoted as $\GG$.
The graphical image for $\GG$ is obtained by deleting the
lines and vertices belonging to  $\Gamma$  from the diagram $G$.
We have already encountered a similar situation
in~\Sect{rf.Definition}, where $\GG$ decomposed into a set of 1PI
\UV-subgraphs. Consider the connected components of $\GG$,
denoting them generically as $\xi$:
\be
   \GG = \prod_i {\xi_i}.
\ee
Denote the set of the loop momenta that are internal with respect to
$\xi$ as $p_\xi$   ($\xi$  may have no loops at all in which case
$p_\xi$  is  empty;  this has no effect on our formalism). Then the
variable $p$ can be decomposed as follows:
\be
     p = (p_\Gamma, p_1, \ldots, p_i, \ldots).
\ee
Then
\be
     \GG(p_{\GG}, M, m, p_\Gamma) = \prod_i \xi_i(p_i, M, m_i),
\ee
where we have introduced the notation $m_i$  for  the  collection
of parameters which contains those light parameters from $m$, as
well as those components  of $p_\Gamma$, on which $\xi_i$  depends.

The \tAs-expansion we are dealing with is independent of the choice of
the operation $\Rm$. in particular, we may fix $\Rm$ to be factorized
in the sense of section~\cref{11} of\1. Then the associated
operation $\cR$ will be also factorized
(see Appendix~\ref{App.Factorization}).
Therefore, performing integration over $p$,
we obtain instead of
\eq{General.Formula} the following expression:
\be[Factorized.As]
     \Asm \. \cR \. \int dp\, G(p,M,m)
     =
     \sumx {\emptyset\leq\Gamma<G} {$\Gamma$ is $m$-subgraph}
     \cR^m \. \int dp_\Gamma \,
	\Bigl(
           \prod_i \Delta^{\rm as} \.\xi_i( M, m_i)
	\Bigr)
        \, \Gamma(p_\Gamma, m),
\ee
where:
\be[Factorized.As.Counterterm]
     \Delta^{\rm as} \.\xi_i( M, m_i)
     =
     \cR \. \int dp_i \,
     \R^m \. \T_{m_i} \. \xi_i(p_i, M, m_i).
\ee
(Note that only the ``counterterms" $\Delta$
are sensitive to the
operation $\cR$ in the initial diagram, while the operation $\cR^m$ used to
perform \tUV\  subtractions on the r.h.s.\ of \eq{Factorized.As} is associated
(in the sense of~\subsect{Structure.Defs}) with the operation $\Rm$ used to subtract
\tIR\  singularities from the formal expansion on the r.h.s.\ of
\eq{Factorized.As.Counterterm}. Recall that $p_i$ are loop momenta of $\xi_i$.)

\paragraph*{A ``fool-proof" recipe for enumeration of subgraphs in
\protect\eq{Factorized.As}.}
It is interesting to note, following\cite{II}, that the condition of
$m$-completeness of $\Gamma$ admits a universal and very convenient
``fool-proof" reformulation.
The above formulae will remain correct if, instead of summing over
$m$-subgraphs $\Gamma$,
one performs summations over all collections of pairwise non-intersecting
and otherwise arbitrary subgraphs $\xi$. Then in order to nullify
superfluos terms it is sufficient to demand that
\textindent{({\it i\/})}
whenever the operation $\Tm$ generates meaningless expressions of the type
$1/0$ (due to a propagator carrying only a combination of light external
momenta) $\Delta_i$ should be put equal to zero;
\textindent{({\it ii\/})} if setting $M=\infty$ in $\Gamma$ produces a factor $1/\infty$ in denominator
(due to a heavy line that happened to remain outside all $\xi_i$'s)
then such terms should be put to zero in the sum in \eq{Factorized.As}.

Such a reformulation is very convenient for studying exponentiation of
expansions of Green functions.

Interpreted graphically, eq.\eq{Factorized.As}
corresponds to shrinking the subgraphs $\xi_i$ to vertices to which
there correspond the factors \eq{Factorized.As.Counterterm} which are
polynomials of the momenta entering the new vertices.

The two formulae \eq{Factorized.As} and
\eq{Factorized.As.Counterterm} represent a fundamental explicit
expression for the \tAs-operation on a renormalized Feynman
diagram.

\subsection{Exponentiation of the \tAs-operation into an OPE-like form.}

The two expressions \eq{Factorized.As} and \eq{Factorized.As.Counterterm}
have exactly the same combinatorial structure as that of the
\tAs-expansions in the dimensionally regularized form studied in\cite{II}.
As was pointed out there, similarity of their structure to that of the
$R$-operation in the MS scheme allows one to easily obtain expansions for
entire collections of diagrams corresponding to Green functions
in the global OPE-like form. In fact, the situation here is even simpler
than in\cite{II} because now all the terms in the expansion
\eq{Factorized.As} which is a starting point for studying
exponentiation, are finite. Therefore, one need not perform the step of
inversion of the $R$-operation, which was the most cumbersome part of\cite{II}
(the role of inverted $R$-operation is played by the operation $\Rm$ in
\eq{Factorized.As.Counterterm}).
Repeating the reasoning of\cite{II}
{\em mutatis mutandis\/} one can immediately write down exponentiated forms
for expansions of Green functions.

Recall that for each asymptotic regime one only
has to find, starting from the basic definitions of the $m$-subgraph $\Gamma$,
diagrammatic characterization of the connected components $\xi_i$ of its
complement $\GG$---wherein the above
``fool-proof" enumeration recipe is very convenient.

\textindent{({\it i\/})} Consider the case corresponding to the short-distance OPE. Then one has
only two (one independent---after taking into account momentum conservation)
heavy external momenta, while all the masses are considered as light
parameters. One finds:
\be[OPE]
     \cR\.\langle T\bigl\{
         \int dx e^{iqx}\, A(x)B(0) \, \exp i[ {\cal L} + \varphi J ]
     \bigr\}\rangle_0
\ee\be
     \asy{q^2}{-\infty}
     \sum_i C_i(q)\, \cR^m\.\langle T\bigl\{
          O_i(0) \, \exp i[ {\cal L} + \varphi J ]
     \bigr\}\rangle_0,
\ee
with the coefficient functions $C_i$ specified by the following
expression:
\be[OPEci]
     \sum_i C_i(q)\, O_i(0)
     =
     \cR\.\Rm\.\Tm\.\int dx e^{iqx}\,
     T\bigl\{
          A(x)B(0) \, \exp i {\cal L}
     \bigr\}^{\rm conn}.
\ee
To correctly interpret these expressions one should keep in mind that
the standard perturbative formalism of interaction representation is used here.
Thus, $A$, $B$ and $O_i$ are local monomials of free fields
({\em without\/} normal ordering) while radiative corrections are generated by
the chronological ($T$-) exponents of the interaction Lagrangian $\cal L$
(integration over the space time is included into $\cal L$).
\tUV\  renormalization is performed by the operations $\cR^m$ and $\cR$.
The operation $\Tm$ acts as follows: one expands the $T$-product
on the r.h.s.\ of \eq{OPEci} in Wick
normal products of the free fields, retains only
connected diagrams that cannot be divided into disconnected parts by
cutting any one of propagators corresponding to the light fields
(cf.\ the above
``fool-proof" recipe)
and the operation
$\Tm$ expands the resulting loop diagrams both in masses and
the momenta corresponding to the free fields in normal products.

Individual coefficient functions
$C_i$ can be extracted by taking corresponding matrix elements of
both sides
of \eq{OPEci} (only tree-level diagrams will be present on the l.h.s.\
since there is no $T$-exponentiated Lagrangian there to generate radiative
corrections).
Such a procedure is analogous to the algorithm of calculating coefficient
functions of OPE in the MS scheme described in\cite{algorithm}.

We see that our formulae are in direct correspondence with
the formulae and algorithms developed
in\cite{fvt:83},\cite{algorithm}, \cite{I},\cite{Inew},\cite{II}.
This should be no wonder because the methods we use were
developed from the very beginning as
a straightforward formalization of the reasoning of those papers.

It is also worth stressing that our formalism contains nothing similar to
the oversubtraction techniques of\cite{zimm} (see also \cite{Tkachov:Advanced} for a discussion of this point).

\textindent{({\it ii\/})} As a second example,
consider the case when the set of heavy parameters $M$ consists of only
heavy masses. Consider the generating functional of Green functions of light
particles:
\be[hm]
     \cR\.\langle T\,\exp[i{\cal L}(\varphi,\Phi)+\varphi J ]\rangle_0,
\ee
where ${\cal L}(\varphi,\Phi)$ is the total
(integrated over space-time) Lagrangian of the system which
depends on heavy and light particles. Supposing that typical momenta
of $\varphi$ and the masses of the light fields $m$ are of the same order of
magnitude and much less than the masses $M$ of the heavy fields, one obtains:
\be[hmex]
     \hbox{eq.\eq{hm}}
     =
     \cR^m\.\langle T\,\exp[i {\cal L}_{\rm eff}(\varphi)
         +\varphi J ]\rangle_0,
\ee
where the effective low-energy Lagrangian whose expression is similar to
\eq{OPEci}:
\be[hmL]
     i{\cal L}_{\rm eff}(\varphi)
     =
     \cR\.\Rm\.\Tm\.\bigl\{
          T\exp i {\cal L} - 1
     \bigr\}^{\rm light-1PI}
\ee\be
     \equiv \sum_n g_{n,\rm eff}(M)\int dx O_n(x),
\ee
where $g_{n,\rm eff}(M)$ are the couplings of the effective Lagrangian.
Note (cf.\cite{II}) that ${\cal L}_{\rm eff}$ can contain
contributions that are quadratic in the light fields. This corresponds to
the finite $M$-dependent field renormalization in the usual formulation of the
decoupling theorem (for a review see\cite{Collins:Ren}.
Also note that only analytic dependence on
the light masses is allowed in ${\cal L}_{\rm eff}$.
``light-1PI" means (cf.\ the above
``fool-proof" recipe) that one has to take
into account only diagrams that have no heavy external fields and such that
they cannot be divided into two disconnected pieces by cutting a line
corresponding to a light particle.

This completes our discussion of the structure of Euclidean asymptotic
expansions of Feynman diagram within the formalism of the \tAs-operation.

\section{Double \tAs-expansions: existence and properties.}
\label{Double}

The aim of the present  section  is  to  prove  a  theorem  on
double \tAs-expansions which summarizes  all  analytic facts
that are  necessary for derivation  of  Euclidean asymptotic
expansions of renormalized Feynman  diagrams.
Explicit formulae have already been presented in section~\sect{Structure}.

We will perform the reasoning in an abstract manner of\1,
without explicit mentioning of Feynman diagrams proper.
All the analytic formulae here---however cumbersome in
appearance---are based on a primordially primitive power
counting. The apparent abstruseness is due to presence of two expansion
parameters and another one used to describe singularities---each of the three
accompanied by an integer-valued index etc. Nevertheless,
the powerful formalism of\1 allows one
to use the recursive structures inherent in the problem and
cut through all the complexities of the proof in an explicit
algebraic fashion.

An abstract mathematical character of the following text
makes it necessary
to recycle some of the physics-inspired notations
used in the preceding sections:
the symbols $m$ and $M$---alongside of $n$ and $N$---will be used
for interger-valued indices while the two expansion parameters will be denoted
as $\k$ and $\s$. We will not need $M$ in its old meaning.
Other notations follow\1.

\subsection{Double \tAs-expansions.}
\label{Double.Formal}

We start with a formal definition of double
\tAs-expansion and present a simple lemma, which
can be considered as a generalization of the uniqueness
property to the case of double \tAs-expansions.

Let $G(\k, \s)$ be a distribution which depends parametrically
on two real parameters $\k$ and $\s$ from
a rectangle $(0,\k_0)\times(0,\s_0)$.

Suppose there exist asymptotic expansions of $G$
in powers and logs of the parameters $\k$ and $\s$.
In the notations of\1 the sum  of terms of
order $\k^n$ is denoted as $\ask^n\.G$ and the partial sum
of the terms through the power $n$,
as $\Ask^n\.G$ (and similarly for $\s$).
By the definition of \tAs-expansion the following asymptotic
estimate must be fulfilled for all $\s$:
\be
     (1 - \Ask^n) \. G = o(\k^n).
\ee
Each term of the expansion is a distribution parametrically depending
on $\s$. Assume that there exist \tAs-expansions of those
distributions
in $\s$. Denote the double series thus obtained as $\Ass\.\Ask\.G$.
We can reverse the order of expansions
and ask a natural question, whether the two resulting double series
$\Ass\.\Ask\.G$ and $\Ask\.\Ass\.G$ coincide.
Generally speaking, they don't
(the simplest example is the numeric function
$1/(\k+\s)$).
However, it is possible to formulate a necessary condition
for the commutativity of the two \tAs-operations
based on the notion of {\em double \tAs-expansion\/}.

Consider the {\em double remainder\/}
     $\ang{(1-\Ask^n)\.(1-\Ass^m)\.G, \varphi}$.
By definition, it is $o(\k^n)$ for all $\s$,
but its behaviour as $\s\to0$ is a priori unpredictable.
It is natural to introduce the following definition:

{\bf Definition.}
A double series in powers and logs of
$\k$ and $\s$---its partial sum of terms through $O(\k^n)\times O(\s^m)$
is denoted as $\Asks^{n,m}\.G$---is called
{\em double \tAs-expansion\/} if:
\be
     \absang{(1 -\Ask^n -\Ass^m +\Asks^{n,m})\.G , \varphi }
     < o(\k^n) \, o(\s^m),
\ee
and there exist integers $n_0$, $m_0$ such that $\asks^{n,m}\.G=0$,
provided $n \le n_0$ or $m \le m_0$ .

One can see that the double \tAs-expansion is unique.
Moreover, its existence implies nice properties
of the double series obtained by termwise
composition of the two one-parameter
\tAs-expansions like $\Ask\.\Ass\.G$, which can be summarized in the
following elementary lemma:

\paragraph*{Lemma 1.}
If there exists a double \tAs-expansion of the graph $G$ then
there exist series $\Ask\.\Ass\.G$ and $\Ass\.\Ask\.G$; moreover:
\be
     \Ass \. \Ask \. G = \Ask \. \Ass \. G = \Asks \. G.
\ee

Therefore, to prove the commutativity of the two \tAs-operations it
is sufficient to construct the double \tAs-expansion---which is the purpose of
the rest of this section.

\subsection{Object of expansion.}
\label{qwer}

The objects we are working with are the graphs,
formally defined in sections~\cref{4} and~\cref{18} of\1.
A graph in this sense is
an abstraction  to describe products
of singular functions encountered on a regular basis
in studies of multiloop  diagrams (e.g.\ integrands
of multiloop diagrams in momentum-representation).
As we now wish to study expansions in two parameters,
the notations of\1 should  be extended.
Namely, the linear functions  $l_g(p)$
(which describe
the  way  the integration (loop) momenta $p$ are combined in the
argument of the $g$-th factor)
is now required to have the form:
\be[def:l]
     l_g(p)
     =
     l'_g(p) + \k \s l''_g + \k l'''_g,
\ee
where $l''_g$ and $l'''_g$ represent linear combinations of  small  and
large external momenta, respectively, and are independent of $p$.
Some of the functions $F_g$
which used to  depend on the expansion parameter $\k$, now acquire
dependence on the second  expansion parameter $\s$ of the form:
\be[def:Fg]
    F_g (q,\k)  \to  F_g (q,\k\s),
\ee
i.e.\ instead of $\k$ we now have the product $\k\s$.
Otherwise, the  properties  of  the functions $F$ remain the same.

The assumptions \eq{def:l}--\eq{def:Fg}  are crucial for existence of double
\tAs-expansion.

To simplify formulae, we assume that the
formal expansions of $F$'s start from $\k^0\s^0=1$,
which can always be achieved by multiplying $F$ by the corresponding
powers of $\k$ and $\s$.

We are going to show that the graph $G(p,\k,\s)$
has a double asymptotic expansion
in powers and logarithms of $\k$ and $\s$ with  the
remainder bounded by an expression in which the dependences on $\k$ and
$\s$ are factorized, i.e.\ that there exist double \tAs-expansion of $G$.

The  theorem  on  double  asymptotic  expansions  presented   below
is, essentially, a logical outcome of the conditions \eq{def:l}--\eq{def:Fg}.
Heuristically, it is clear why this is so: a numeric function of the
form
$1/(1+\k+\k\s)$
with  a structure analogous to \eq{def:l}, can be expanded into a
double \tAs-expansion in $\k$  and $\s$. The latter property is
naturally inherited  by  any products  of  such functions.  The
pathological cases of the sort mentioned in~\subsect{Double.Formal}
are prohibited by the imposed restrictions.

\paragraph*{Theorem 1.}
Under the above conditions, there exists a double asymptotic
expansion of the graph $G( p, \k, \s)$; it is given by
a termwise composition of the two \tAs-operations as described in the preceding
sections (see also below eqs.\eq{asks:def}--\eq{E:def};
and it has the following properties:

\textindent{({\it a\/})} $\Ask\.\Ass\.G = \Ass \. \Ask \. G = \Asks\.G $.

\textindent{({\it b\/})}
$\Asks^{n,m}\.G=0$ \ for \ $n<A^{\k G}$ \ or \ $m<A^{\s G}$, where
\be
     A^{\k G}
     =
     \mathop { \mathop {\max}_{\Gamma\leq G} }
     _{\hbox{\scriptsize $\Gamma$ is $\k$-subgraph}}
     (0, \o_\Gamma),
     \quad\quad\quad\quad
     A^{\s G}
     =
     \mathop { \mathop {\max}_{\Gamma\leq G} }
     _{\hbox{\scriptsize $\Gamma$ is $\s$-subgraph}}
     (0, \o_\Gamma).
\ee

\textindent{({\it c\/})} The operation $\Asks$ is local in the sense of\1.

\textindent{({\it d\/})} For the terms of the expansion the following estimate holds:
for all $\varphi\in\D(P)$ such that ${\rm rad\,supp\,}\varphi\leq d$,
\be
     \absang{ \asks^{n,m}\.G , \varphi }
     \leq
     \k^n\s^m d^{-\o^\0_\G-n} \S[\varphi,d] \L(\k,\s).
\ee
(Here and below we do not indicate the upper limits of summations
since their exact expressions are cumbersome and of no practical use.
They can, however, be restored from the proofs in a straightforward manner.)

\textindent{({\it e\/})} The remainder of the double expansion, defined as
\be
     \Delta_{n,m} \bydef {1-\Ask^n-\Ass^m+\Asks^{n,m}},
\ee
satisfies the following ``factorizable" estimate:
one can fix a constant  $C>0$ such that for all $d>C\k$
and $\varphi \in\ D(P)$ with ${\rm rad\,supp\,}\varphi \leq  d$:
\be[Th:(e)]
     \absang{ \Delta_{n,m}\.G , \varphi }
     \leq
     \k^{n+1}\s^{m+1}d^{-\o^\0_\G-n-1}
     \S[\varphi,d] \L(\k,\s).
\ee

\textindent{({\it f\/})}
The expansion possesses the following {\em minimality property\/}:
\be
       \ang{ \asks^{n,m}\.G * \P_{a,G} }
     = \ang{ \Delta_   {n,m}\.G * \P_{a,G} }
     = 0,
     \quad\hbox{for}
     \quad\abs{a}\le\o^\0_\G+n.
\ee

\subsection{Proof of theorem 1\EOH.}

The proof of the theorem will be carried out by induction
over the hierarchy of $m$-subgraphs $\Gamma<G$. It is convenient
to include into the induction the following lemma containing useful
auxiliary estimates:

\paragraph*{Lemma 2.}
\begin{itemize}
\item[({\it i\/})] $\forall\varphi\in\D(\P),{\rm\ \ rad\,supp\,}\varphi\leq d$:
\be
     \absang{\ask^n\.(1-\Ass^m)\.G}
     \leq
     \k^{n}\s^{m+1} d^{-\o^\0_\G-n}
     \S[\varphi,d] \L(\k,\s),
\ee
\item[({\it ii\/})]
$\exists C>0\;\,\forall d>C\k\;\,
\forall\varphi\in\D(P),\quad {\rm rad\,supp\,}\varphi\leq  d $:
\be
     \absang{ \ass^m\. {(1-\Ask^n)}\.G}
     \leq
     \k^{n+1}\s^{m} d^{-\o^\0_\G-n-1}
     \S[\varphi,d] \L(\k,\s).
\ee
\end{itemize}

The statements of the theorem and the lemma are trivial for the empty
graph $G=1$. Let us suppose that they hold for any subgraph of
$G$. The proof can be divided into three logical steps.
First, one
defines \tAs-expansion as a distribution on $\D(P\backslash\{0\})$
using decompositions of unit (cf.\  section~\cref{21} of\1)
and verify the
conditions of the theorem for it.
Second, one performs a natural extension
of the distributions obtained at the first step onto the space of
functions from $\D(P)$ with zero of the order $B_n=\o_\G+n+1$
(such space is denoted  as $\D_{B_n}(P)$).
Third, one continues the \tAs-expansion onto the entire $\D(P)$
and determines a finite renormalization to ensure asymptotic estimates.

It is convenient to carry out the first and second steps
simultaneously.

\paragraph*{Steps 1--2.}
To begin with, take a function $\varphi\in\D_{B_n}(P)$ and
a cutoff $\cut_\l\in\D(P\backslash\{0\})$. Using the sector
decomposition of unit we define $\asks\.G$ on $\varphi\cut_\l$:
\be
     \ang{ \asks^{N,M}\.G \,,\, \varphi\cut_\l }
     =
     \sum_{ \Gamma\triangleleft G}
     \mathop{\sum_{ n\leq N}}
                 _{ m\leq M}
     \ang{ \asks^{n,m}\.\Gamma \,
         , \, \tk^{N-n}\.\ts^{M-m}\.\GG \,
           \theta_\Gamma \, \varphi \cut_\l
         }.
\ee
Using the estimate ({\it d\/}) for $\Gamma<G$ (which holds by inductive assumptions)
and the auxiliary estimate \eq{tt:est} we conclude that:
\be
     \absang{ \asks^{N,M}\. G \,,\, \varphi\cut_\l }
     \leq
     \k^n \s^m
     \sum_{k\geq B_n}
        \l^{k-\o^\0_\G-n}
        \mathopen\Vert\varphi\mathclose\Vert^k \L(\l)\,\L(\k,\s).
\ee
Integration over $\l$ (cf.\ sect.\cref{21} of\1) completes
steps 1--2 for the estimate ({\it d\/}).

The estimate ({\it i\/}) is proved similarly using the auxiliary estimate \eq{tD:est}.

The estimates ({\it ii\/}) and ({\it e\/}) are of a somewhat different kind,
which should be clear from their look.
First of all, we prove the following lemma:

\paragraph*{Lemma 3.}
$\forall\varphi,\;\; {\rm rad\,supp\,}\varphi\leq C\k$:
\be
     \absang{ \ass^m\. G \,,\, \varphi }
     \leq
     \s^m \k^{-\o^\0_\G} \S[\varphi,\k] \L(\s),
\ee
\be
     \absang{ (1-\Ass^m) \. G \,,\, \varphi }
     \leq
     \s^{m+1} \k^{-\o^\0_\G} \S[\varphi,\k] \L(\s).
\ee

The unusual way of how $\k$ enters the r.h.s.\ is due to two reasons.
First, now the test function is localized in a neighbourhood of radius
$O(\k)$ so that $\k$ plays the role normally reserved for $d$.
Second, the formal expansion of $G$ in $\s$ prior to expansion in $\k$
results in a situation with several maximal subgraphs (in the context of this
lemma we are dealing only with $m$-subgraphs). This would normally
prevent one from obtaining estimates describing singular behaviour of subgraphs
(recall that in the proofs of\1 one normally deals with
one maximal subgraph whose singular plane is the point $p=0$ so that
behaviour
near $p=0$ can be described by dependence on the radius of support of
test functions). In the present case, however,
the singular product expanded in $\s$ depends parametrically on
$\k$---in an interesting way (here the reader
should review the pattern of how the factors $G$ depend
on $\k$ and $\s$---see~\subsect{qwer}).
Consider the factors that develop singularities
after expansion in $\s$ but prior to expansion in $\k$.
The only dependence on $\k$ that remains in such factors is in their
momentum arguments. This means that the eventual expansion in $\k$ will
result in an $O(\k)$ shift\footnote{
and/or rotation in the more general situation
of the expansion problem with contact terms.
}
of their singular planes,
while after expansion in $\k$ there will remain only one maximal subgraph
($G$ itself) whose singular plane is $p=0$. Therefore, our standard estimates
are still meaningful provided the test functions satisfy the
condition in the lemma.

The proof proceds as follows.
In section~\cref{8} of\1 there was constructed a decomposition
of unit isolating the singular planes of maximal
($m$-) subgraphs of $G$, whereby a smooth function $\cut_\G$ is assigned
to each $\Gamma\in\Smax[G]$, so that
\be[decunit]
     \sum_{\Gamma\in\Smax[G]} \cut_\Gamma(p) \equiv 1.
\ee
It is convenient (in fact, natural) to choose $\cut_\Gamma$ to have
the form $\cut_\Gamma(p/\k)$, so that the decomposition of unit works for all $\k$.
Using \eq{decunit}, one can write down the following identity:
\be
     \ang{ \ass^m\. G \,,\, \varphi}
     =
     \sum_{ \Gamma\in\Smax[G]} \,\, \sum_{ m \leq M }
     \ang{ \ass^m\.\Gamma \,,\, \ts^{M-m} \. \GG\,
                            \cut_\Gamma \,
                            \varphi
         }.
\ee

Since every $\Gamma$ is maximal in itself, the parameter $\s$
appears in its expansion only in combination $\k\s$.
Therefore, the estimate (\cref{20.10}) from\1 may be
applied to  $\ass^m \. \Gamma$ after replacing $d\to\k$
and $\k \to \k\s$:
\be
     \absang{ \ass^m\. G \,,\, \varphi }
     \leq
     \mathop{\max}_{ \Gamma\in\Smax[G] }
     \,
     \mathop{\max}_{ m\leq M }
     \,
     (\k\s)^m
          \k^{-m-\o^\0_\Gamma}
          \S\bigl[ \ts^{M-m}\.\GG\,
              \cut_\Gamma
              \varphi,
              \k
            \bigr] \L(\s).
\ee
(We assume that $\s\le 1$ and, hence, $\s\k\le\k$
and $\k>{\rm rad\,supp\,}\varphi$.)
Noticing that:
\be
     \bigl\Vert \ts^{M-m} \. \GG \bigr\Vert
     ^k
     _{ {\rm supp} (\cut_\Gamma\varphi) }
     \leq
     \s^{M-m} \k^{-d^\0_\GG-k} \L(\s, \k),
\ee
we obtain the desired estimate.
The second statement of the lemma is proved similarly.

Now let us return to the theorem.
We now have to deal with singularities of the expansion in $\k$ and, therefore,
with $\k$-subgraphs.
To prove the estimates ({\it ii\/}) and ({\it e\/}) we choose the decomposition
of unit $1 = \Cut^\k(p) + \Cut_\k(p)$
($\Cut^\k$ non-zero in a neighbourhood of $p=0$),
with $\Cut^\k$ fixed so that $\forall \Gamma\triangleleft G$
and $\forall g\in\GG$:
\be
     {\rm supp} (\Cut_\k\theta_\Gamma) \cap {\cal O}^\k_{g(G)} = \emptyset,
\ee
where $ {\cal O}^\k_{g(G)}$ is the $\k$-vicinity of $g$ defined
as in\1 but with the following modifications.
The singular plane  $\pi_g^\s$ of an element $g$ at $\s=0$ can be
displaced from the one at $\k=0$, $\pi_g^\k$, by ${\rm const}\times\k$.
Therefore we can take a neighbourhood of the plane
$\pi_g^\k$ with a radius ${\rm const}\times\k$ and
containing the singular plane $\pi_g^\s$ for all $\k\to0$.

Since ${\rm rad\,supp\,}\Cut^\k=C\k$,
we use our usual representation of $\Cut_\k$ as an integral over spherical layers
of radius $\l$, $\Cut_\k=\int_{C\k}^d\,d\l/\l\,\cut_\l$,
and get for ({\it ii\/}) with $\varphi$ replaced by $\varphi \cut_\l$
in analogy with the proof of ({\it d\/}),({\it i\/})
for $\l > C\k$ (using  \eq{Dt:est}):
\be
     \absang{ \ass^m \. (1-\Ask^n)\. G \,,\, \varphi\cut_\l }
     \leq
     \k^{n+1}\s^m
     \sum_{k\ge B_n}
          \l^{k-\o^\0_\G-n-1}
          \mathopen\Vert\varphi\mathclose\Vert^k \L(\l)\,
          \L(\k,\s).
\ee
Integrating over $\l$ from  $C\k$ to $d$  we immediately
obtain a ``half" of ({\it ii\/}) (i.e.\ for $\varphi \Cut_\k$ instead of $\varphi$).
The second half,
\be
     \ang{ \ass^m \. (1-\Ask^n) \. G \,,\, \varphi \Cut^\k }
     =
     \ang{ \ass^m \. G \,,\, \varphi \Cut^\k }
     -
     \sum_{ l \leq m }
         \ang{ \ass^l \. \ask^n \. G \,,\, \varphi \Cut^\k },
\ee
is estimated with the help of lemma 3 (the first term)
and ({\it d\/}) which has been already proved.
The estimate ({\it e\/}) may be obtained in the same manner.
This completes steps 1--2 of the proof.

\paragraph*{Step 3.}
The extension procedure have already been performed for
the series $\Ask\.G$ and $\Ass\.G$ in\1.
Our purpose is to extend $\Asks\.G$ to the distribution
on $\D(P)$ while preserving all the estimates.
It can be done in two steps.
First, one applies
the special subtraction operator $\r$ to $\asks^{n,m}\. G$ so
that it obey the estimates ({\it d\/}). This procedure is performed
similarly to the reasoning of\1. Second one adjusts
the finite renormalization of $\asks^{n,m}\.G$
so that it stisfies ({\it d\/})
(and, of course, ({\it i\/}) and ({\it ii\/})).
Namely, let:
\be[asks:def]
     \asks^{n,m} \. G
     =
     \r \. \as^{n,m}_{\k,\s} \. G
     +
     \sum_{\abs{a}=\max(\o_\G+n,0)}
     \d_a(p) \, E^{\k,\s}_{ a,m}.
\ee
It is straightforward to check that the choice
\be[E:def]
     E^{\k,\s}_{ a,m}
     =
     \ang{ \P_{a,G} *
	 \bigl[
              \ass^m \. G
            - {\rAs}^n \. \ass^m \. G
	 \bigr]
     }
\ee
satisfies all the requirements, which completes the proof of the theorem 1.

A few remarks are in order.
\textindent{({\it a\/})}
The  equations \eq{asks:def} and \eq{E:def} give explicit
recursive formulae for the double \tAs-expansion.
They can be resolved along
lines of~\Sect{rf.Definition} which
has already been done in the preceding section.
\textindent{({\it b\/})}
The theorem can be readily generalized
to the case of $N$-fold expansions. For example, if one wished
to study a two-scale expansion of a renormalized diagram,
one would have to use a triple \tAs-expansion etc.
\textindent{({\it c\/})}
In our main inequality \eq{Th:(e)} the graph $G$
is compared to a rather weird expression $\Ask\. G+\Ass\. G-\Asks\. G$.
However, this is only necessary for obtaining factorized bounds.
Should one need just an approximation for $G$ irrespective of whether
it should be achieved due to smallness of $\k$ or $\s$ or
both---as is often the case in
applications---it is sufficient to use a suitably truncated series
$\Asks\. G$.

\section*{Conclusions\EOH.}

A theory that claims to be a comprehensive alternative to the BPHZ method
should be able to address at the formal level, as a minimum,
the problem of \tUV\  renormalization and  that of
operator product expansions.
The theory of \tAs-operation,
which has already enjoyed success in applications,
has now fulfilled this criterion.

As was observed in\cite{paradigm},\cite{Tkachov:Advanced}
the key  difference  between  the two paradigms---BPHZ and our
techniques based on the \tAs-operation---is how
the basic dilemma  of  the theory of multiloop
Feynman diagrams is  resolved. The dilemma consists in the conflict
between the inherently recursive
nature of the problems of perturbative quantum field theory involving
hierarchies of Feynman diagrams, and the singular nature of the objects
participating in such recursions. The  BPHZ approach consists in
systematically getting rid of the singularities by
explicitly resolving the corresponding recursive patterns and thus
reducing the problem to absolutely convergent integrals.
However, those recursive structures are inherently natural,
and to ignore them---as the BPHZ approach does---means to lose
the heuristic advantage of dealing with complicated objects
in a manner respectful of their true nature.
Moreover, in the case of non-Euclidean asymptotic regimes,
the corresponding recursions are much more complex
and can hardly be resolved in an explicit form\cite{Tkachov:Advanced}.

The technique of the \tAs-operation,
on  the contrary, allows one to preserve  and  make
efficient use  of  the  recursive structures by offering means to
directly work with singular expressions.
The key anlytical idea of the theory of the \tAs-operation
is to use the locality condition which explicates
the recursive strucures, and thus to reduce
the problem to studying the singularities localized
at an isolated point.
As a result, formal proofs become
algebraically explicit and compact, while the final calculational formulae,
powerful.

To put it shortly: the BPHZ formalism is only an instrument of formal proof;
the techniques of the \tAs-operation is also an instrument of discovery.

\subsection*{Acknowledgements.}

The authors are grateful to D.~Robaschik, A.~A.~Slavnov, D.~A.~Slavnov
and G.~'t~Hooft for stimulating discussions.
We thank  V.~V.~Vlasov for help
in clarifying some technical points.
The work of one of the authors (A.~K.)
is supported in part by the Weingart Foundation
through a cooperative agreement with the Department of Physics at UCLA.
The other author (F.~T.) thanks the
CTEQ collaboration for support and J.~C.~Collins
for discussions of extension of the techniques of \tAs-operation
to Minkowskian asymptotic regimes, and NIKHEF, FERMILAB and Penn State,
where parts of this work were done, for hospitality.
This work was supported in part by the Russian Foundation for Basic Research under grant 95-02-05794.

\newpage

\begin{appendices}

\appendix{Factorizability of the operation $\protect\cR$\EOH.}
\label{App.Factorization}

Let us prove  factorizability  of  the  operation $\cR$.
It is always possible to fix the associated $\tilde{R}$-operation
to be factorized whence the
factorization of $\cR$ follows.
For clarity,
we consider the case of two factors.

Let $G'(p', \k)$ and $G''(p'', \k)$ be two graphs, with $p'$
and  $p''$ independent.  We assume that the operation $\As_\k$  is
well-defined on both of them. We wish  to prove that,
provided the operation $\r$ is chosen to be factorized,
the operation
$\cR=1-\rAs$
factorizes as follows:
\be[B.1]
     \cR \. \int dp'\,dp''\, G'(p', \k) \, G''(p'', \k)
     =
     \cR \. \int dp'\, G'(p', \k)
     \times
     \cR \. \int d p''\, G''( p'', \k).
\ee
We will present  simple  arguments  which use only factorizability
of  the operation $\As_\k$ (which follows from uniqueness of
\tAs-expansions---see~\subsect{As.Uniqueness} of\1)
and the fact that the expression
$\cR \. \int dp \, G(p)$ (where $p=(p',p'')$)
is exactly  the  coefficient  of $\d(p)$ in
the \tAs-expansion of $G(p, \k)$ in $\k$ in the sense of distributions:
\be[B.2]
     \As_\k \. G(p,\k)  = \rAs\.G(p,\k)
     + \d(p) \, \cR \. \int d\bar p\,G(\bar p,\k)
     + \hbox{higher derivatives of $\d(p)$}.
\ee
The proof runs as follows. First one writes:
\be[B.3]
     \As_\k \. \bigl[G'(p',\k) \times G''(p'',\k)\bigr] \kern7cm
\ee\be
     = \rAs \. \bigl[G'(p',\k) \times G''(p'',\k)\bigr]
     + \d( p') \d( p'')
       \, \cR \. \int d \bar p'd \bar p''\, G'(\bar p',\k) G''(\bar p'',\k)
     + \ldots.
\ee
Then for each factor one has a similar expression; e.g.\ for $G'$:
\be[B.4]
     \As_k\.G'(p',\k) = \rAs\.G'(p',\k)
     + \d( p') \,\cR \. \int d\bar p'\,G'(\bar p',\k)
\ee
Recall that $\As_k$ factorizes (see~\SUB\cref{15.9} of\1):
\be[B.5]
     \As_\k\. \bigl[G'(p',\k) G''(p'',\k)\bigr]
     = \As_\k\. G'(p',\k) \times \As_\k\. G''(p'',\k).
\ee
Substituting \eq{B.3} and \eq{B.4} into \eq{B.5}  and  comparing
terms (taking into account that we always choose $\r$ to be factorizable)
one finds, first, that
\be[B.6]
     \rAs\. \bigl[G'(p',\k) \times G''(p'',\k)\bigr]
     =
     \rAs\. G'(p',\k) \times \rAs\.G''G''(p'',\k)
\ee\be
     + \As_\k\. \Gamma'(p',\k) \times \d( p'')\,
       \cR \. \int d \bar p''\,\Gamma''(\bar p'',\k)
     + \As_\k\. \Gamma''(p'',\k) \times \d( p')\,
       \cR \. \int d \bar p'\,\Gamma'(\bar p',\k),
\ee
whence follows \eq{B.1}.

It remains to note that the factorization conditions that we always impose
on $\r$  and  $\R$ ensure factorizability of $\cR$.

\appendix{Auxiliary estimates\EOH.}
\label{App.Aux.Est}

Let $H$ be a subproduct of $G$, $H\subset\G$,
and let $K$ be a compact region not intersecting any of the singular planes of
$H$, which is formally expressed as follows:
\be
     K \subset P_{(G)} \backslash \cup_{g\in H} \pi_g.
\ee
Then
\be[tt:est]
     \bigl\Vert \tk^n \. \ts^m \. H \bigr\Vert
     ^k
     _{ p \in \l K }
     <
     \left( {\k \over \l } \right)^n \s^m \,
     { \L(\l, \k) \over \l^{d_H+k} } ,
\ee
\be[tD:est]
     \bigl\Vert \tk^n \. (1-\Ts^m) \. H \bigr\Vert
     ^k
     _{ p \in \l K }
     <
     \left(    { \k \over \l }
     \right)^n
     \s^{m+1} \,
     { \L(\l, \k) \over \l^{d_H+k} }.
\ee
Moreover, there exists a constant $C$ (depending on $H$ and $K$)
such that for $\l>C\k$:
\be[Dt:est]
     \bigl\Vert (1-\Tk^n) \. \ts^m \. H \bigr\Vert
     ^k
     _{ p \in \l K }
     <
     \left( { \k \over \l }
     \right)^{n+1} \s^m \,
     { \L(\l, \k) \over \l^{d_H+k} },
\ee
\be[DD:est]
     \bigl\Vert (1-\Tk^n) \. (1-\Ts^m) \. H \bigr\Vert
     ^k
     _{ p \in \l K }
     <
     \left( { \k \over \l } \right)^{n+1} \s^{m+1}
     { \L(\l, \k) \over \l^{d_H+k} }.
\ee
All the above estimates formalize elementary power counting with respect to
each of the three parameters---$\k$, $\s$ and $\l$.

\end{appendices}

\newpage

\section*{Figure Captions\EOH.}

\bigtextindent{{Fig.1.}} An example of the \IR-subgraph is shown
with fat lines in ({\it a\/}). One can take $p_1$ and $p_2$ for
its proper variables $p_\gamma$. ({\it b}) corresponds to the
``projection'' $\G\backslash\gamma$.

\medskip

\bigtextindent{{Fig.2.}}
({\it a\/}) is the graph $\G$, the fat lines constitute one
of its \IR-subgraphs $\gamma$.
({\it c\/}) is the diagrammatic representation for $\gamma$
considered on its own (recall that we consider integrands prior to any loop
integrations), while
({\it b\/}) corresponds to $[\G\backslash\gamma]_\gamma$.

\medskip

\bigtextindent{{Fig.3.}}
An example of a general \IR-subgraph $\gamma$ (shown with fat lines).
Its proper variables are $p_\gamma=(p_1,p_2,p_3)$,
while $p^{\rm int}_\gamma=p_2$ and
$p^{\rm ext}_\gamma=(p_1,p_3)$.
The complement $\G\backslash\gamma$  (formed by thin lines)
falls into two \UV-subgraphs $\xi_1$ and $\xi_2$
whose loop momenta are $p_{\xi_1}=p_4$ and $p_{\xi_2}=p_5$.

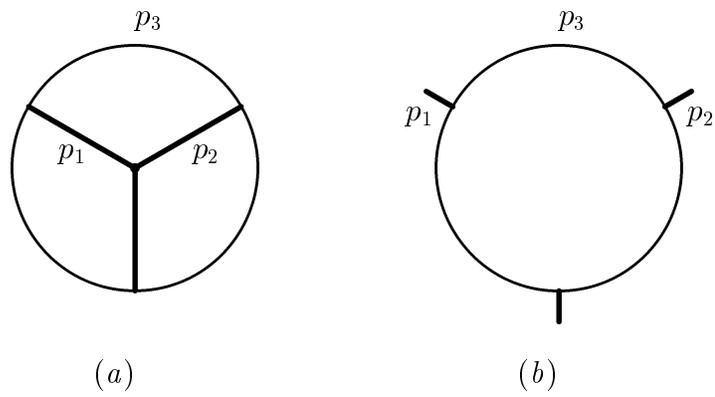
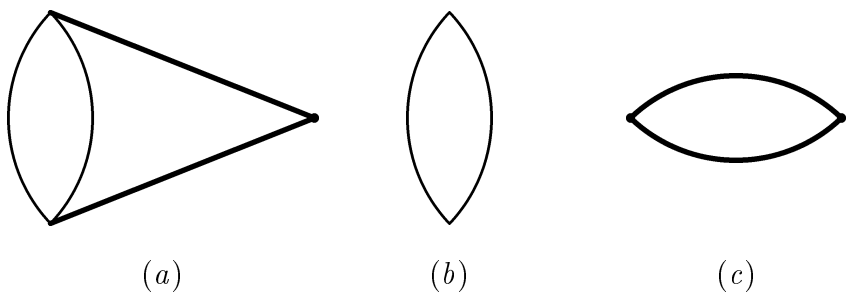
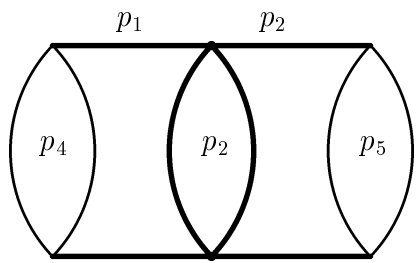
\end{document}